\documentclass[journal]{IEEEtran}
\usepackage{epsfig,cite}
\usepackage{graphicx}
\graphicspath{{figures/}} 
\usepackage{color}
\usepackage{amssymb,amsmath,mathtools, mathrsfs}
\usepackage{enumitem}
\usepackage{graphicx}
\usepackage{epstopdf}
\usepackage[ruled,vlined]{algorithm2e}

\definecolor{blue}{rgb}{0,0,1}
\definecolor{darkgreen}{rgb}{0,.5,0}
\definecolor{darkred}{rgb}{.5,0,0}

\newtheorem{theorem}{Theorem}
\newtheorem{proposition}{Proposition}

\newtheorem{rem}{Remark}

\def\levy{L\'evy }

\def\jth{$j^{\text{th}}$}

\def\BinomDist{{\mathscr{B}}}
\def\PoissDist{{\mathscr{P}}}

\DeclareMathOperator\erfc{erfc}
\DeclareMathOperator\erfcinv{erfcinv}

\IEEEoverridecommandlockouts

\begin{document}
	\bstctlcite{IEEEexample:BSTcontrol}
	
	\title{Capacities and Optimal Input Distributions \\for Particle-Intensity Channels}
	
	\author{Nariman Farsad\IEEEauthorrefmark{1}
    Will Chuang \IEEEauthorrefmark{5}, 
    Andrea~Goldsmith\IEEEauthorrefmark{1},
    Christos Komninakis\IEEEauthorrefmark{2},
    Muriel M\'edard\IEEEauthorrefmark{3},\\
    Christopher~Rose\IEEEauthorrefmark{4},
    Lieven Vandenberghe\IEEEauthorrefmark{5},
    Emily E. Wesel\IEEEauthorrefmark{6}, and
    Richard D. Wesel\IEEEauthorrefmark{5}\thanks{ This research is supported in part by National Science Foundation (NSF) grant CCF-1911166 and by the NSF Center for Science of Information under Grant CCF-0939370. Any opinions, findings, and conclusions or recommendations expressed in this material are those of the author(s) and do not necessarily reflect the views of the NSF. This paper was presented in part at ISIT 2017 \cite{ParticleIntensityModulation} and the 2018 ITA Workshop \cite{richard_d._wesel_efficient_2018}.}
        \thanks{\IEEEauthorrefmark{1}Nariman Farsad and Andrea Goldsmith are with the Department of Electrical Engineering, Stanford University, Stanford, CA, 94305. Emails: \{nfarsad, andreag\}@stanford.edu.}
        \thanks{\IEEEauthorrefmark{2}Christos Komninakis is with Qualcomm, 6455 Lusk Blvd, San Diego, CA, 92121.  Email:  christos@qti.qualcomm.com.}
        \thanks{\IEEEauthorrefmark{3}Muriel M\'edard is with the Department of Electrical Engineering and Computer Science, MIT, Cambridge, MA, 02139.  Email:  medard@mit.edu.}
        \thanks{\IEEEauthorrefmark{4}Christopher~Rose is with the School of Engineering, Brown University, Providence, RI, 02912. Email: christopher\_Rose@brown.edu.}
        \thanks{\IEEEauthorrefmark{5}Will Chuang, Lieven Vandenberghe, and Richard D. Wesel are with the Department of Electrical and Computer Engineering, UCLA, Los Angeles, CA 90095-1594. Emails: \{chuangw, vandenbe, wesel\}@ucla.edu.}
        \thanks{\IEEEauthorrefmark{6}Emily E. Wesel is with Stanford University, Stanford, CA, 94305, Email: ewesel@stanford.edu.}
	}

	\maketitle

	\newcommand{\note}[1]{{}}
	\newcommand{\notenf}[1]{{}}
	
	\begin{abstract}
		This work introduces the particle-intensity channel (PIC) as a model for molecular communication systems and characterizes the capacity limits as well as properties of the optimal (capacity-achieving) input distributions for such channels. In the PIC, the transmitter encodes information, in symbols of a given duration, based on the probability of particle release, and the receiver detects and decodes the message based on the number of particles detected during the symbol interval. In this channel, the transmitter may be unable to control precisely the probability of particle release, and the receiver may not detect all the particles that arrive. We model this channel using a generalization of the  binomial channel and show that the capacity-achieving input distribution for this channel always has mass points at probabilities of particle release of zero and one. To find the capacity-achieving input distributions, we develop an efficient algorithm we call dynamic assignment Blahut-Arimoto (DAB). For diffusive particle transport, we also derive the conditions under which the input with two mass points is capacity-achieving. 
	\end{abstract}
	

	\section{Introduction}
	
	In molecular communication (MC) transmitters convey information by releasing small particles.  Information may be contained in the number or type of released particles or in the time of release \cite{far16ST}. These particles travel to the receiver where they are detected and the message decoded. The stochastic nature of the transport process introduces uncertainty about the time of particle release and even the number of particles released during a given symbol interval.

	One approach to understanding the capacity limits of molecular channels investigated in prior work assumes information is encoded in the time instants at which particle(s) are released. Such channels are called molecular timing channels (MTCs). In particular, the additive inverse Gaussian noise channel is presented in \cite{sri12,li14}, and upper and lower bounds on capacity are derived. These works assume a system where information is encoded in the release time of a {\em single} particle.  Molecular timing channels where information is encoded via the release times of {\em multiple} particles are considered in \cite{rose2016inscribed}, which presents upper and lower bounds on capacity, and \cite{far18_MTCcap} introduces a MTC where particles decay after a finite interval and derives upper and lower bounds on the associated capacity.	
    
	Another approach to MC encodes information through the number of particles released at the transmitter and decodes based on the number of particles that arrive at the receiver during the symbol interval. We focus on this type of modulation scheme and call it {\em particle-intensity modulation} (PIM)\footnote{This has been called the concentration-shift-keying or the amplitude-modulation in previous work. However, we believe PIM captures the physical properties of this system and its relation to optical intensity modulation.}.

	In \cite{ein2011, tah15}, this concentration-based channel is considered with a receiver equipped with ligand receptors. The process of molecule reception of a ligand receptor is modeled as a Markov chain and the capacity in bits per channel use is analyzed. The results are extended to multiple access channels in \cite{ami15PtoP}. In \cite{gha15}, a binomial distribution is used to model a system where the transmitter can perfectly control the release of particles and the receiver can perfectly detect the number of particles that arrive.  It is assumed that the channel has finite memory and particle transport is assisted by flow. Using this model, bounds on the capacity are derived, and the capacity for different memory lengths is analyzed. Reference \cite{ami15} assumes that the channel input is the rate of particle release. The channel is represented as a Poisson channel with finite memory, and upper and lower bounds on capacity per channel use are presented. Finally, different channel coding schemes are compared for MC systems that employ PIM in \cite{Lu15}.

	This paper extends our conference papers in \cite{ParticleIntensityModulation,richard_d._wesel_efficient_2018} where we considered molecular channels with imperfect PIM and imperfect detection. We call this channel the {\em particle-intensity channel~(PIC)}. Specifically, in the PIC the sender releases particles independently and probabilistically (i.e., information is encoded in the release probability), and the destination may not detect all the particles that arrive. Note that this is a different formulation than \cite{ParticleIntensityModulation} in that the channel input is continuous as opposed to discrete. We assume that the duration of the symbol is long enough that particles from one symbol have a negligible effect on future symbols. This model is reasonable if particles diffuse beyond the receptor or disappear in some other fashion, for instance through degradation \cite{guo16}. Under this assumption, the PIC is memoryless. Finally, we assume that particles can be generated at a constant fixed rate at the transmitter. 
	
	For this model, we show that the PIC can be represented with a channel model similar to the binomial channel \cite{KomninakisISIT2001} where the input is the probability of success and the output is the number of successes in a fixed number of Bernoulli trials.  Like the binomial channel, the PIC channel input (probability of particle release) is continuous over the interval $[0, 1]$.  However, unlike the original binomial channel, the probability of success (i.e., of the particle being detected by the receiver) is smaller than channel input so that the maximum probability of success is less than 1.  This introduces asymmetry in the behavior induced by the extreme channel inputs; a zero induces a deterministic result at the receiver but a one does not. 
	
	Another difference from the original binomial channel is the introduction of a symbol duration.  The number of trials in the PIC, which is the maximum number of particles that can be released by the transmitter, changes as a function of symbol duration because particles are generated at a constant rate at the transmitter. 

	Our contributions in this work are as follows.
\begin{itemize}
	\item This paper defines the capacity of the PIC channel in bits per second and as a function of symbol duration. We show that this channel is related to the binomial channel. To the best of our knowledge this is the first time that a channel model for molecular communication is presented that includes imperfections at {\em both} transmitter and receiver.
	\item This paper demonstrates that the optimal input distribution for the PIC channel, for any symbol duration, always has mass points at the two extremes 0 and $1$. We also derive an expression for the capacity when the input is binary, and present the conditions under which binary input achieves capacity. 
    \item This paper presents the {\em dynamic assignment Blahut-Arimoto} (DAB) algorithm as a new algorithm for finding the capacity and the optimal finite-support input distribution for many channels with continuous input alphabets, including the PIC channel.  This algorithm converges much faster than the ellipsoid method and finds the minimum-cardinality capacity-achieving input distribution.  Using DAB, this paper calculates the capacities and minimum-cardinality capacity-achieving input distributions for a wide range of channel parameters.
    \item Although the binary input distribution (i.e., on-off-keying) is capacity achieving for a large class of PICs based on diffusive particle transport, this paper shows that the capacity-achieving input distribution has more than two mass points when the probability of particle arrival is sufficiently high.
\end{itemize}	
	
	The rest of this paper is organized as follows. In Section \ref{sec:model} we present the PIC. Then in Section \ref{sec:ChanCapOptInp}, we formulate the capacity, investigate characteristics of the optimal input distribution, and derive the capacity of the binary input PIC. Section \ref{sec:ellipsoid} presents the ellipsoid method for finding the optimal input distribution for the binomial channel, while in Section \ref{sec:DAB}, the DAB algorithm is developed. We present numerical results in Sections \ref{sec:results}, and in Section \ref{sec:conclusion} we discuss the concluding remarks.      

	\section{The Particle Intensity Channel (PIC)}
	
	\label{sec:model}
	
The PIC is an MC channel in which information is communicated through PIM, i.e., the channel input  $X$ is the probability of particle release by the transmitter. The transmitter controls the intensity of the released particles by controlling this probability. The particles themselves are assumed to be {\em identical and indistinguishable} at the receiver, and no other properties (such as the time-of-release) are used for encoding information. The receiver then counts the number of particles that arrive during the symbol duration to produce the channel output $Y$. The particles released by the transmitter travel to the receiver through a propagation mechanism (e.g., diffusion). We assume that the particles {\em travel independently of each other}, and are {\em detected independently of each other}. This is a reasonable model used in many previous works \cite{far16ST}. 	

According to the PIC model, particles are released {\em instantly and simultaneously} at the beginning of the symbol interval.  Particles are released independently of each other, and the transmitter's intended probability of release is $X$, which is the channel input. The transmitter controls the number of particles that are released by changing $X$, e.g., by controlling the degree of opening in a nozzle or porous membrane. Therefore, the channel input is continuous over the interval $X \in [0,1]$. 

\begin{figure}
\vspace{-0.1cm}
\begin{center}
\includegraphics[width=0.75\columnwidth,keepaspectratio]{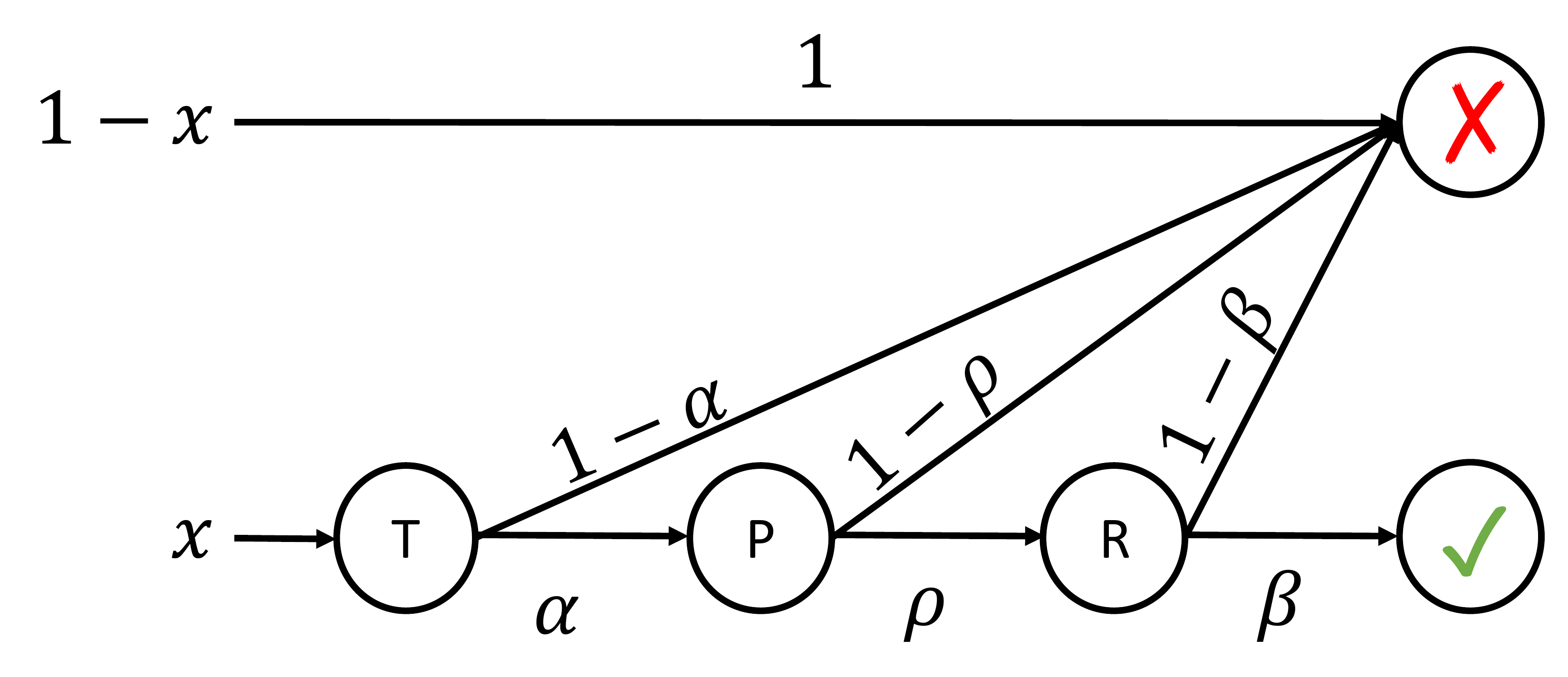}
\end{center}
\vspace{-0.35cm}
\caption{\label{fig:PIchannel} The stochastic transmission (T), propagation (P) and reception (R) of a single particle.  At the transmitter, particles are selected to be released with probability $x$.  A selected particle is actually released with probability $\alpha$.  Through propagation, a released particle arrives at the receiver within the symbol duration with probability $\rho$.  A particle that arrives at the receiver is detected with probability $\beta$.  Thus, a particle selected for release is detected at the receiver with probability $\theta_{\rho} = \alpha \rho \beta$.}
\vspace{-0.5cm}
\end{figure}

Figure \ref{fig:PIchannel} shows the stochastic release, arrival, and detection of a single particle.  We define the probability of release failure to be $1-\alpha$ where $0<\alpha \leq 1$ .  Thus the actual probability that each particle is released is $\alpha x$, when the transmitter wants to transmit with probability $x$. 

We now consider the stochastic particle transport. Each released particle will arrive at the receiver at some independent identically distributed random time $T \sim f_T(\cdot)$.  Let $f_T(t)$ denote the PDF of the time the particle arrives, and $F_T(t)$ denote its corresponding CDF. Then the probability that a released particle arrives during a symbol duration $\tau$ is given by	
	\begin{align}
	\label{eq:DefRho}
	\rho = F_T(\tau),
	\end{align}
and the probability that it never arrives (and is by assumption never detected) is $1-\rho$.

In the PIC model, each particle that arrives at the receiver is detected with probability $\beta$.  This detection process is i.i.d.  In this way, the PIC model  incorporates the receiver's  inability to perfectly detect the particles that arrive, owing to sensitivity or uncertainty in the detection process.  

Thus, as shown in Figure \ref{fig:PIchannel} a particle selected for release is detected at the receiver with probability $\theta_{\rho} = \alpha \rho \beta$.
Let $Y$ be the number of particles that are actually detected during the corresponding symbol duration.  The end-to-end channel between the input $X$ and the output $Y$ is the binomial $\BinomDist(m_{\rho},x\theta_{\rho})$:
\begin{align}
\label{eq:PIChanTxRx}
P(Y=y|X=x) = \binom{m_\rho}{y} (x \theta_{\rho})^{y} (1-x \theta_{\rho})^{m_\rho-y},
\end{align}
where $\BinomDist(n,p)$ indicates the binomial distribution with parameters $n$ and $p$, and $m_\rho$ is the number of particles available for release, which we now discuss.

The channel is used in a time-slotted fashion, where $\tau$ is the symbol duration. We define a parameter $\lambda$ as a constant fixed rate at which the transmitter can generate particles. We assume that $m_\tau = \left\lfloor \lambda\tau \right\rfloor$ particles are available to be released by the transmitter at the beginning of each time slot, and the transmitter releases each with channel input probability $X$. Note that in this model, the number of particles that can be released at the beginning of each time slot can change with the symbol duration $\tau$.

If we assume $F_T(t)$ is strictly monotone, the symbol duration $\tau$ can be obtained from $\rho$ by using the inverse CDF (iCDF) function, i.e., $\tau=F^{-1}_T(\rho)$. Using the iCDF, we can also rewrite $m_\tau$ as a function of $\rho$:
	\begin{align}
	m_\rho  = \left \lfloor \lambda F^{-1}_T (\rho) \right \rfloor \, .
	\end{align}

Particles that neither  arrive at the receiver nor dissipate over the symbol duration could interfere with detection during future channel uses. Such intersymbol interference (ISI) should be incorporated into deriving the channel capacity.  While such ISI represents an interesting area for future investigation, in this paper we assume that particles with transit times exceeding $\tau$ dissipate or are otherwise inactivated.  That is, particles are assumed to have a finite lifetime of duration $\tau$. This assumption seems reasonable since particles could be rendered undetectable either naturally or by design (via denaturing or gettering/enzyme reactions \cite{guo16}). Under this assumption, channel uses are independent and the maximum mutual information between input and output during a single channel use defines the channel capacity.

An important observation here is that $\rho$ and $m_{\rho}$ change as a function of the symbol duration $\tau$. In this work, we incorporate the optimization of the symbol duration into the formulation of capacity to determine the channel capacity of the memoryless PIC in bits per second. This is one important distinction between this and previous work such as \cite{sri12,li14,gha15,ami15}, where the channel capacity is typically defined in bits per channel use.
	
For the case when $m_\rho$ is large and $\theta_{\rho}$ is small, the system can be well approximated by the Poisson distribution \cite{ver69}
\begin{align}
\label{eq:PIChanPois}
P(Y=y|X=x) = \frac{(x \theta_\rho m_\rho)^{y}e^{x \theta_\rho m_\rho}}{y!}.
\end{align}
We write this as $P(y|x;\rho) \sim \PoissDist(x \theta_\rho m_\rho)$, where $\PoissDist(a)$ indicates the Poisson distribution with parameter $a$. 
	
	\begin{rem}
		Using the Poisson approximation, the PIC in MC systems can be viewed as a more general formulation of the discrete-time Poisson channel used to model optical intensity channels \cite{sha90,lap98,cao14}. Because a finite number of particles are released, particle arrival rate does not increase linearly with time and thus neither does the capacity. This is in contrast to the discrete-time Poisson channel in optical communications, where photon arrival rate increases linearly with the symbol duration \cite{cao14}. PIC symbol durations that are too long can reduce the information rate, as demonstrated in Section \ref{sec:results}. Note that although we do not consider interfering particles, they can be introduced to the Poisson model in \eqref{eq:PIChanPois} by adding an extra term similar to the dark current in optical communications \cite{sha90,lap98,cao14}. 
	\end{rem}
	
	\section{Channel Capacity and Optimal Input}

	\label{sec:ChanCapOptInp}

	We now characterize the channel capacity of the PIC. Let $f_X(x)$ be the channel input PDF and let $\mathcal{F}$ be the set of all valid input PDFs. Then the capacity of the channel in \eqref{eq:PIChanTxRx} as a function of the particle arrival probability $\rho$ and having units of [bits per second] is defined as
	\begin{align}
	\label{eq:CapPIasP}
	\mathsf{C}(\rho) =  \underset{f_X(x)\in\mathcal{F}}{\max}  \frac{I(X;Y|\rho)}{F_T^{-1}(\rho)},
	\end{align}
	where $F_T^{-1}(\cdot)$	is the iCDF of  the particle detection time. 
	Since the channel changes as a function of the symbol duration, the fundamental limit of this channel is then
	\begin{align}
	\label{eq:capPI}
	\mathsf{C}^*=\underset{\rho}{\max}~\mathsf{C}(\rho).
	\end{align}
	
	
	We now investigate the characteristics of the optimal input distribution, determine capacity under a binary input constraint, and investigate settings for which the capacity-achieving distribution is binary. First, observe that although the input distribution is over a finite interval, the capacity-achieving input distribution has finite support, requiring at most $m_\rho+1$ mass points for the PIC in \eqref{eq:PIChanTxRx}.  This was proven in \cite{WitsenhausenIT1980} using Dubin's theorem \cite{Dubins1962}.  See also \cite{GallagerBook1968} (Corollary 3 in Chapter 4.5).
	
	Throughout the paper we will use a tilde over a letter to indicate that this is an ordered set or vector whose cardinality is equal to the number of mass points being used by the associated input distribution.  Let $\mathcal{\tilde{X}}=\{x_0, x_1, \cdots, x_{m_\rho}\}$ be the location of the $m_\rho+1$ mass points with $0 \leq x_0 < x_1 < \cdots < x_{m_\rho}\leq 1$. Let $\mathcal{\tilde{P}} = \{p_{x_0}, p_{x_1},\cdots,p_{x_{m_\rho}}\}$ be the probabilities corresponding to each mass point. Then the corresponding input distribution is given by
    \begin{align}
        \label{eq:optInputForm}
    	f_X(x)=\sum_{i=0}^{m_\rho} p_{x_i}\delta(x-x_i). 
    \end{align}
The optimal input distribution always has the form of \eqref{eq:optInputForm}. We now show that the optimal input always has non-zero mass points at the two extremes of  $X=0$ and $X=1$.
	
	\begin{theorem}
		\label{thm:massPoints}
		For a given symbol duration $\tau$, and hence a given $\rho$, the mutual information given in \eqref{eq:CapPIasP} is maximized by a PDF $f_X^*(x)$, where $p^*_0>0$ and $p^*_1>0$. 
	\end{theorem}
	
	\begin{IEEEproof}
		The first inequality $p^*_0>0$ can be proved by using \cite[Lemma 1]{cao14} and the second inequality $p^*_1>0$ can be proved using \cite[Lemma 3]{cao14}.
	\end{IEEEproof}	
	
	We now derive the capacity in \eqref{eq:CapPIasP} as a function of $\rho$ for binary input PIC (i.e., a system that is limited to on-off-keying). Note that conveniently for the PIC in \eqref{eq:PIChanTxRx}, on-off-keying is equivalent to $X=1$ or $X=0$.

	\begin{theorem}
		\label{thm:binaryCap}
		Let  $\tilde{\cal X}_b = \{0,1\}$ be the selected input alphabet for the PIC in \eqref{eq:PIChanTxRx} with $p_1 = P(X=1)$, and $\varphi_\rho=(1-\theta_\rho)^{m_\rho}$. The optimal input distribution $p_1$ under this binary input constraint is given by 
		\begin{align}
		\label{eq:binOptiInput}
		p_1^*=\frac{1}{\varphi_\rho^{\tfrac{\varphi_\rho}{\varphi_\rho-1}}-\varphi_\rho+1}, 
		\end{align} 
		and the capacity of \eqref{eq:CapPIasP}, in bits per second, is given by
		\begin{align}
		\label{eq:binCapCEMperP}
		\mathsf{C}^b(\rho)= \frac{1}{F_T^{-1}(\rho)} \log\left( 1+(1-\varphi_\rho)\varphi_\rho^{\tfrac{\varphi_\rho}{1-\varphi_\rho}}\right).
		\end{align}	  
	\end{theorem}
	
\begin{IEEEproof}
	Define $Y^+$ as the indicator function
	\begin{equation}
	   Y^+ = \mathbf{1}(Y>0).
	\end{equation}
	$Y^+$ is a sufficient statistic of $Y$ for $X$ \cite{cover-book} so that the binary-input PIC is equivalent to a Z channel \cite{tal02}.
	Thus, the mutual information in \eqref{eq:CapPIasP} can be written as a function of $p_1$ using
		\begin{align}
		I_{\rho}(X^b;Y) &= I_{\rho}(p_1)\\ 
		\label{eq:mutualBinary}
		&=H(p_1(1-\varphi_\rho)) - p_1H(\varphi_\rho).
		\end{align}
		Setting the derivative of $I_{\rho}(p_1)$ with respect to $p_1$ equal to zero yields \eqref{eq:binOptiInput}. Substituting \eqref{eq:binOptiInput} into \eqref{eq:mutualBinary} and using \eqref{eq:CapPIasP} we obtain the capacity expression in \eqref{eq:binCapCEMperP}. 
	\end{IEEEproof}	

	An interesting question arises here as to when the binary input alphabet $\tilde{\cal X}_b$ is optimal for the PIC. In the following proposition, we provide a guideline for the optimality of the binary input for a  subclass of PICs.
	\begin{proposition}
		\label{prop:CondiBinary}
		For the PIC in \eqref{eq:PIChanTxRx} where $m_\rho$ is large and $\theta_{\rho}$ is small such that the Poisson approximation in \eqref{eq:PIChanPois} is accurate, the binary input distribution given in \eqref{eq:binOptiInput} is optimal if $m_\rho \theta_{\rho} < 3.3679$.
	\end{proposition}
	\begin{IEEEproof}
		Using the same technique presented in \cite{sha90} for the optical channels, the proposition can be proved. 
	\end{IEEEproof}	
	
	Note that this condition may be satisfied in many practical systems where the radius of the receiver is much smaller than the distance between the transmitter and the receiver, hence the probability of particles arriving is small. Upper bounds on the total variation between binomial and Poisson distributions can be used to show that this variation is small for small $\theta_{\rho}$~\cite{ver69}.


\section{Computing Capacity of the Binomial Channel}
\label{sec:ellipsoid}

The previous section provides an expression for the capacity when the optimal input distribution is binary. This section and the next address how to compute the capacity of the PIC channel when the optimal input is not binary. Several papers have addressed similar capacity computations including \cite{KomninakisISIT2001}, \cite{non-uniform}, \cite{poisson}, \cite{poisson_part_1}, \cite{constrain}, \cite{char}, \cite{optical}, \cite{photon}, \cite{poisson_2}, and \cite{poisson_3}.

This section lays the foundation for Section \ref{sec:DAB}  to introduce the DAB algorithm as a general solution technique for computing the capacity and optimal finite-support input distribution for channels with continuous input alphabets. Subsection  \ref{sec:BinomialChannel} introduces the binomial channel, which is a corner case of the PIC channel. Subsection \ref{sec:convex_optimization} formulates the binomial channel capacity problem as a convex optimization problem, and presents its dual.  Subsection \ref{sec:Ellipsoid} explores the Ellipsoid method as one technique to solve the dual problem.


\subsection{The Binomial Channel}
\label{sec:BinomialChannel}
To simplify our initial development, we focus on the corner case of the PIC channel  \eqref{eq:PIChanTxRx} where $\theta_{\rho}=1$.  In this case the PIC channel is the binomial channel  (of parameter $n=m_p$) and  has a channel law defined by the binomial probability distribution of order $n$ \cite{KomninakisISIT2001}.  For each channel use, the input $X$ is the probability of success of a Bernoulli trial.  The channel output $Y$ is the number of successes observed during $n$ Bernoulli trials.  Thus the channel transition probability law is described as 
\begin{equation}
P^{(n)}_{Y|X}(y|x) = \begin{pmatrix}n\\y\end{pmatrix}x^y(1-x)^{n-y} \, , \label{eq:binomial}
\end{equation}
where the possible  $y$ values are the integers zero through $n$.  The channel output could also be the ordered list of Bernoulli trial outcomes, but since $Y$ is a sufficient statistic \cite{CoverBook} of those outcomes for estimating $X$, the capacity is the same.

 \begin{figure}[t]
\centering\includegraphics[width=21pc]{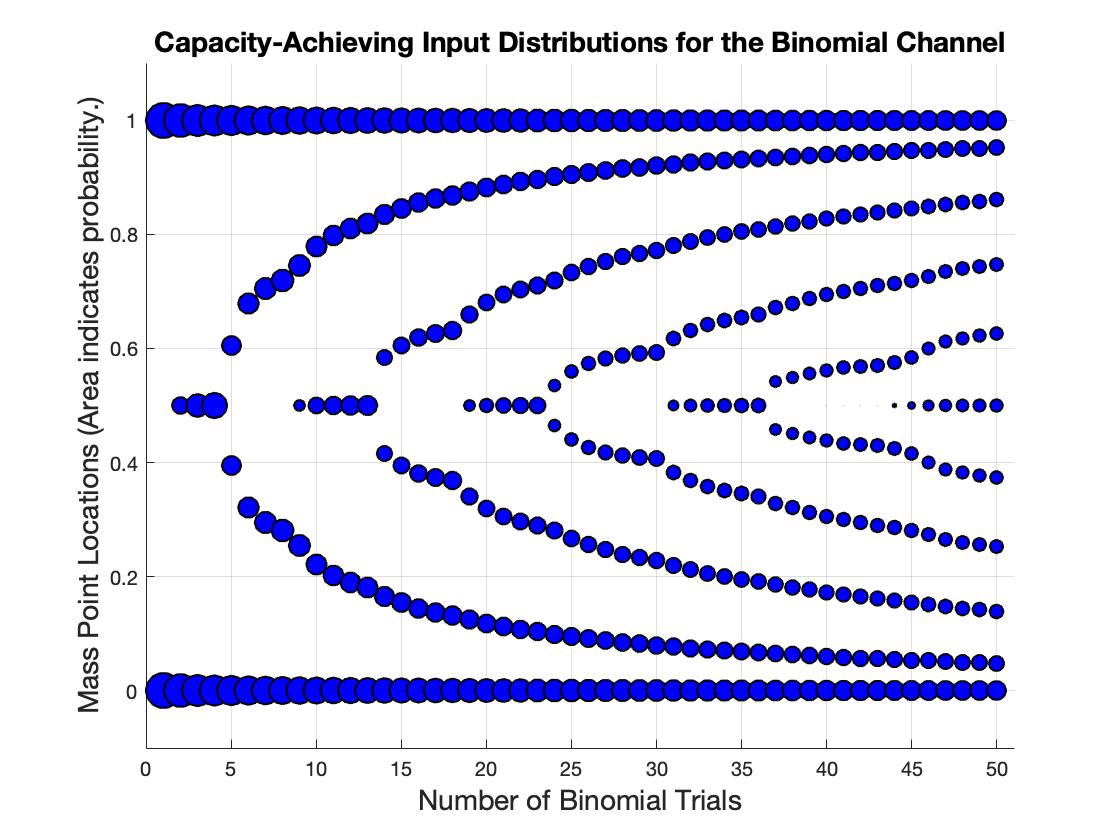}
\caption{Capacity-achieving input distributions for the binomial channel with $1 \le n \le 50$ obtained by the Dynamic Assignment Blahut-Arimoto algorithm described in Section \ref{sec:DAB}.}
\label{Figure:MassPoints}
\end{figure}

 Figure \ref{Figure:MassPoints} shows these finite-support capacity-achieving distributions for $1 \le n \le 50$, with the area of the circle indicating the probability of the mass point.   For $n=1$, the mass points are at zero and one, resulting in a noiseless binary channel.  At $n=2$, a mass point is introduced at 0.5,  growing in probability as $n$ increases, until at $n=5$ that mass point splits into two.  As $n$ increases, these two mass points move away from 0.5, and when, at $n=9$, they are far enough away, a new mass point is born at 0.5.


\subsection{A Convex Optimization Problem and its Dual}
\label{sec:convex_optimization}

Despite the fact that finite-support distributions achieve capacity of the PIC, direct application of the Blahut-Arimoto algorithm \cite{BlahutTIT1972} is complicated because the locations of the support points within the unit interval are not known.  Reasonable approximations can be obtained by applying Blahut-Arimoto with mass points closely spaced along the entire unit interval, with most of these having zero probability.  However, we are interested in algorithms that identify the capacity more precisely and that explicitly identify the mass points of the capacity-achieving distribution.

 In \cite{KomninakisISIT2001}, the capacity of the binomial channel is computed by first formulating the problem as a convex optimization problem and then solving it by using the Ellipsoid method. Assuming that $X$ has discrete support, capacity $C_n$   is
 \begin{align} 
 \small
 C_n &= \max_{ f_X(x)} I(X;Y)\\
 &= H(Y) - H(Y|X)\\
 &= \max_{ f_X(x)}  \left \{ H(Y) - \int_{x=0}^1 f_X(x) H(Y|X=x)  f_X(x) dx \right \} \, . \label{eq:integratedfx}
 \end{align}
Despite the fact that  the capacity-achieving distribution on $X$ has at most $n+1$ mass points, the distribution on $X$ is expressed as a density function  $f_X(x)$  (and an integral is used in \eqref{eq:integratedfx}) because the positions of the support points are located anywhere in the uncountable set of the unit interval.  Thus,  $f_X(x)$ consists of a countable number of delta functions located anywhere in the unit interval.
 
The optimization problem of \eqref{eq:integratedfx} can be formulated as a convex optimization problem in a vector space with uncountably infinite dimension.  For more mathematical precision, replace $f_X(x)dx$ with $dF(x)$ where $F(x)$ is the cumulative distribution.  We allow  $dF(x) \in {\cal F}$, the set of signed measures on the unit interval and include additional constraints to force $dF(x)$ to be a valid probability distribution.  Introducing the additional variables $q_y$ for $y \in \{0, 1, \ldots, n \}$ and appropriate equality constraints that force the $q_y$ values to be the output probability distribution $P_Y(y)$ induced by the input distribution yields the following convex optimization {\bf primal problem}:
\begin{align*}
\text{minimize}~~&\sum_{y=0}^n q_y\log q_y +  \int_{x=0}^1 f_X(x) H(Y|X=x) dF(x)\\
\text{subject to}~~&-dF(x) \le 0, \forall x \in [0,1]\\
&\int_{x=0}^1 dF(x) - 1 = 0 \, ,\\
&q_y -   \int_{x=0}^1 P^{(n)}_{Y|X}(y|x) dF(x) = 0, ~  y \in \{0, \ldots, n \} \, .
\end{align*}

The infinite dimensional $dF(x)$ makes the problem intractable.  We create a Lagrangian dual problem that can be solved with traditional methods.  We introduce Lagrange multipliers $\mu, z_0, z_1, \ldots, z_n$    for the equality constraints and the measurable mapping $v(x)$ of $[0,1]$ to the one-dimensional real space $\bf R$ for the inequality constraint producing Lagrangian $L({\bf  q}, dF(x), v(x), {\bf z}, \mu)$ \cite{Luenberger}:
\begin{align*}
L= &\sum_{y=0}^n q_y\log q_y + \int_{x=0}^1 H(Y|X=x) dF(x)\\
& -   \int_{x=0}^1 v(x)dF(x) + \mu \left( \int_{x=0}^1 dF(x) - 1 \right)\\
& + \sum_{y=0}^n z_y \left( q_y - \int_{x=0}^1 P^{(n)}_{Y|X}(y|x) dF(x) \right) \, ,
\end{align*}
which is the cost function augmented with the weighted sum of the constraints. Minimizing $L({\bf  q}, dF(x), v(x), {\bf z}, \mu)$ with respect to primal variables  $\bf q$ and $dF(x)$  gives the dual function $g(v(x), {\bf z}, \mu)$ as follows:
\begin{equation*}
g=  \inf_{ {\bf q}, dF(x)} \left \{ \sum_{y=0}^n q_y (z_y + \log q_y) - \mu +  \int_{x=0}^1 \gamma(x)dF(x)  \right\},
\end{equation*}
where 
\begin{equation}
 \gamma(x) = H(Y|X=x) -v(x) + \mu -\sum_{y=0}^n z_y P^{(n)}_{Y|X}(y|x) \, .
\end{equation}

Because $dF(x)$ is an unconstrained unsigned measure, $g(v(x), {\bf z}, \mu) = -\infty $ unless $\gamma(x) \geq 0$ for all $x \in [0,1]$, in which case we have
\begin{equation}
g(v(x), {\bf z}, \mu)  = \inf_{\bf q}  \left \{ \sum_{y=0}^n q_y (z_y + \log q_y) - \mu  \right\}   . \label{eq:infq}
\end{equation}

Setting $d/dq_y$ of the summation in \eqref{eq:infq} to zero yields the minimizing value of $q_y =\frac{2^{-z_y}}{e}$ so that
\begin{equation}
g(v(x), {\bf z}, \mu)  =  \frac{- \log e}{e} \sum_{y=0}^n 2^{-z_y} - \mu   \, \label{eq:final_g}.
\end{equation}
The { \bf dual problem} for our primal problem maximizes this $g(v(x), {\bf z}, \mu)$ subject to constraints on the slack variables:
\begin{align*}
\text{maximize}~~& \frac{- \log e}{e} \sum_{y=0}^n 2^{-z_y} - \mu \\
\text{subject to}~~& v(x) \ge 0~~ \forall x \in [0,1],\\
&H(Y|X=x) -v(x) + \mu -\sum_{y=0}^n z_y P^{(n)}_{Y|X}(y|x) =0 \, .
\end{align*}
Combining these two constraints eliminates the cumbersome infinite-dimensional $v(x)$ producing
\begin{align*}
\text{minimize}~~&\sum_{y=0}^n 2^{-z_y} + \frac{\mu e}{\log e}\\
\text{subject to}~~&H(Y|X=x) + \mu -\sum_{y=0}^n z_y P^{(n)}_{Y|X}(y|x)  \ge 0 ~~ \forall x \in [0,1] \, .
\end{align*}
Minimizing the objective function requires the minimum possible value of $\mu$ that satisfies 
\begin{equation}
\mu  \ge \sum_{y=0}^n z_y P^{(n)}_{Y|X}(y|x) -H(Y|X=x)~~ \forall x \in [0,1]\, ,
\end{equation}
which leads to the final formulation of the dual problem, in which only the variables $z_y$ remain:
\begin{equation*}
\min_{\bf z} \sum_{y=0}^n \frac{2^{-z_y}}{e} + \frac{1}{\log e} \max_{x \in [0,1]} \left \{ \sum_{y=0}^n z_y P^{(n)}_{Y|X}(y|x) -H(Y|x) \right\}.
\end{equation*}

The dual problem is a finite variable  convex optimization problem over the vector $\bf z$, which can be solved using a variety of techniques.  

Once the minimizing $\bf z$ vector is obtained, complementary slackness indicates that the capacity-achieving mass points are the $x$ values that maximize $\sum_{y=0}^n z_y P^{(n)}_{Y|X}(y|x) -H(Y|x) $.  The output distribution is recovered using $P_Y(y) = q_y =\frac{2^{-z_y}}{e}$, and the probability $P^*_X(x)$ associated with each mass point can be found by solving the equations:
\begin{equation}
P_Y(y)=\sum_{x \in A} P^{(n)}_{Y|X}(y|x) \mathcal{\tilde{P}}(x) ~~ \forall y \in \{0, \ldots, n\},
\end{equation}
where $A$ is the set of maximizing $x$ values.  The mutual information induced by this  $P_X$ is also the capacity.

 \subsection{The Ellipsoid Method}
 \label{sec:Ellipsoid}

In \cite{KomninakisISIT2001}, the ellipsoid method was used  to solve the dual problem identified in Section \ref{sec:convex_optimization}.  The ellipsoid method was developed by Shor, Nemirovski, and Yudin in the 1970's and used by Khachiyan \cite{Khachiyan1979} in 1979 to show the polynomial solvability of linear programs.  See \cite{EllipsoidSurvey1981} for an excellent survey.  One conclusion of \cite{EllipsoidSurvey1981} is that the ellipsoid method, while of academic interest, is often not the fastest way to solve a convex problem and can have stability issues as well.  However, it is straightforward to program. 

The method begins with an initial ellipsoid ${\cal E}^{(0)} \in \mathbb{R}^N$ centered at a point $z_0$ , which is defined as 
\begin{equation}
{\cal E}^{(0)} = \left \{  z \in \mathbb{R}^N : (z - z_0) ^TP_0^T (z - z_0) \le 1\right \} \, , 
\end{equation}
and is known to contain the optimizing point $z^*$.  At the $k^\text{th}$~iteration, the point $z_k$ is at the center of the ellipsoid
\begin{equation}
{\cal E}^{(k)} = \left \{  z \in \mathbb{R}^N : (z - z_k) ^TP_k^T (z - z_k) \le 1\right \} \, .
\end{equation}
To compute the ${\cal E}^{(k+1)}$ we need the subgradient $g_{k+1} \in \mathbb{R}^N$ which is a vector that satisfies 
$g_{k+1}^{T} (z^*-z_k) \le 0$,  so that 
\begin{equation}
z^* \in {\cal E}^{(k)}  \cap \left \{  z : g_{k+1}^{T} (z-z_k) \le 0 \right \} .
\end{equation}
The following computations create a new ellipsoid that contains the half-ellipsoid described above:
\begin{align}
\tilde{g}_{k+1} &=\left (  \sqrt{g_{k+1}^{T}P_k  g_{k+1}} \right )^{-1} {g}_{k+1}\\
z_{k+1} &= z_k - \frac{1}{N+1} P_k \tilde{g}_{k+1}\\
P_{(k+1)} &= \frac{N^2}{N^2 - 1}\left( P_k - \frac{2}{N+1}P_k \tilde{g}_{k+1}\tilde{g}_{k+1}^{T}P_k \right ) \, .
\end{align}

\begin{figure}
\centering\includegraphics[width=21pc]{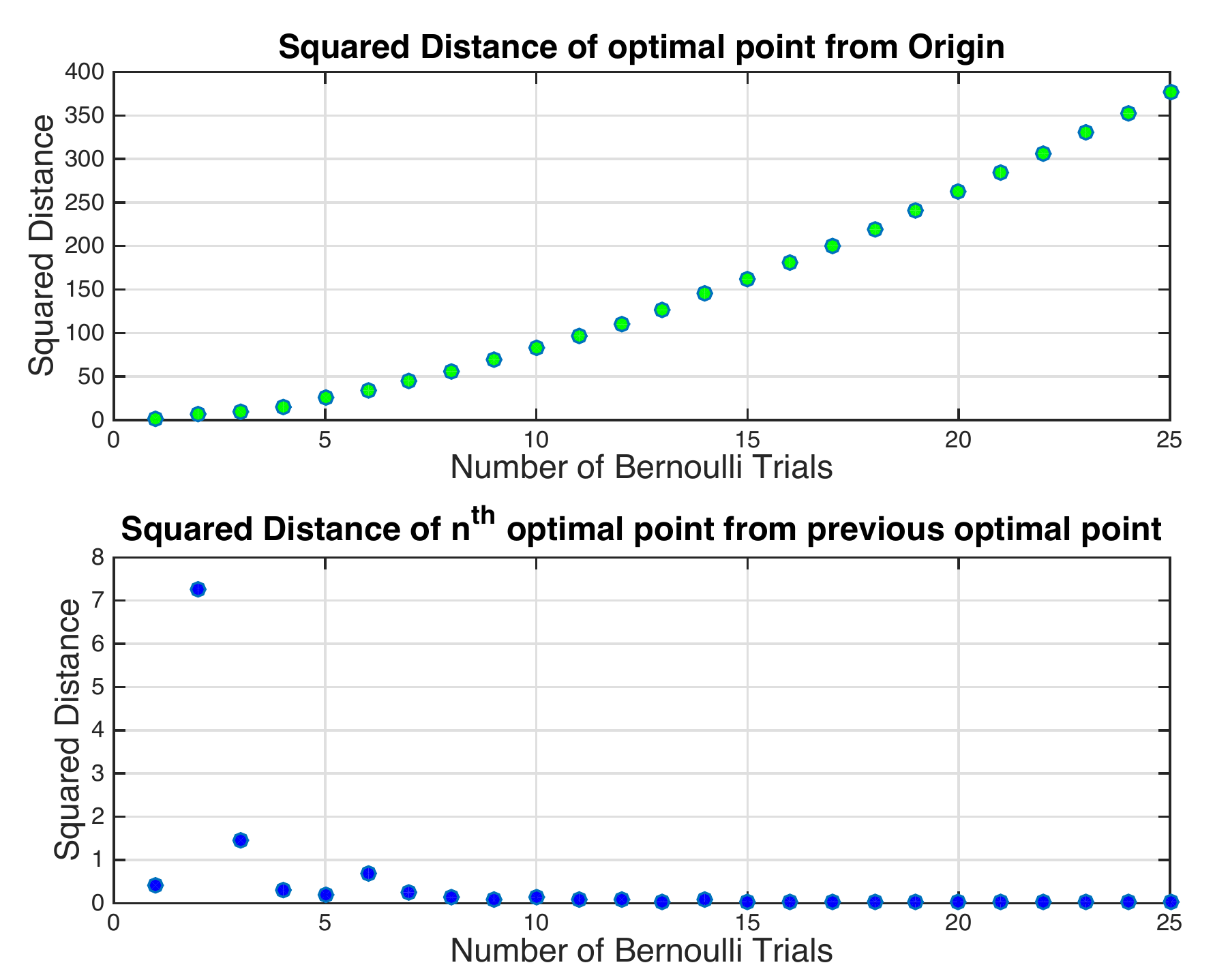}
\caption{Squared distance from the origin of optimal points.}
\label{Figure:distances}
\end{figure}

The ellipsoid method stopping criterion computes $\sqrt{g_k^TP_kg_k}$ which is an upper bound on the error in the objective function.
To apply the ellipsoid algorithm to the dual problem of Section \ref{sec:convex_optimization}, $N=n+1$ and the subgradient is the vector with elements  
\begin{equation}
- \frac{2^{-z_y} - e P_{Y|X}(y|x^*)}{\log e} \, ,
\end{equation}
where $x^*$ is any value of $x$ that maximizes 
\begin{equation}
\sum_{y=0}^n z_y P^{(n)}_{Y|X}(y|x) -H(Y|x)
\end{equation}
for the current set of $z_y$ values.  Also needed are $z_0$ and $P_0$ that create an initial ellipse that contains $z^*$.   The simplest approach is to select the origin for $z_0$ and choose $P_0$ to be the identity  scaled by a value that is larger than the square of the optimizing $z^*$.  Figure \ref{Figure:distances} shows for the first 25 values of $n$ these squared distances, which grow to over 350 by $n=25$.

\begin{figure}
\centering\includegraphics[width=21pc]{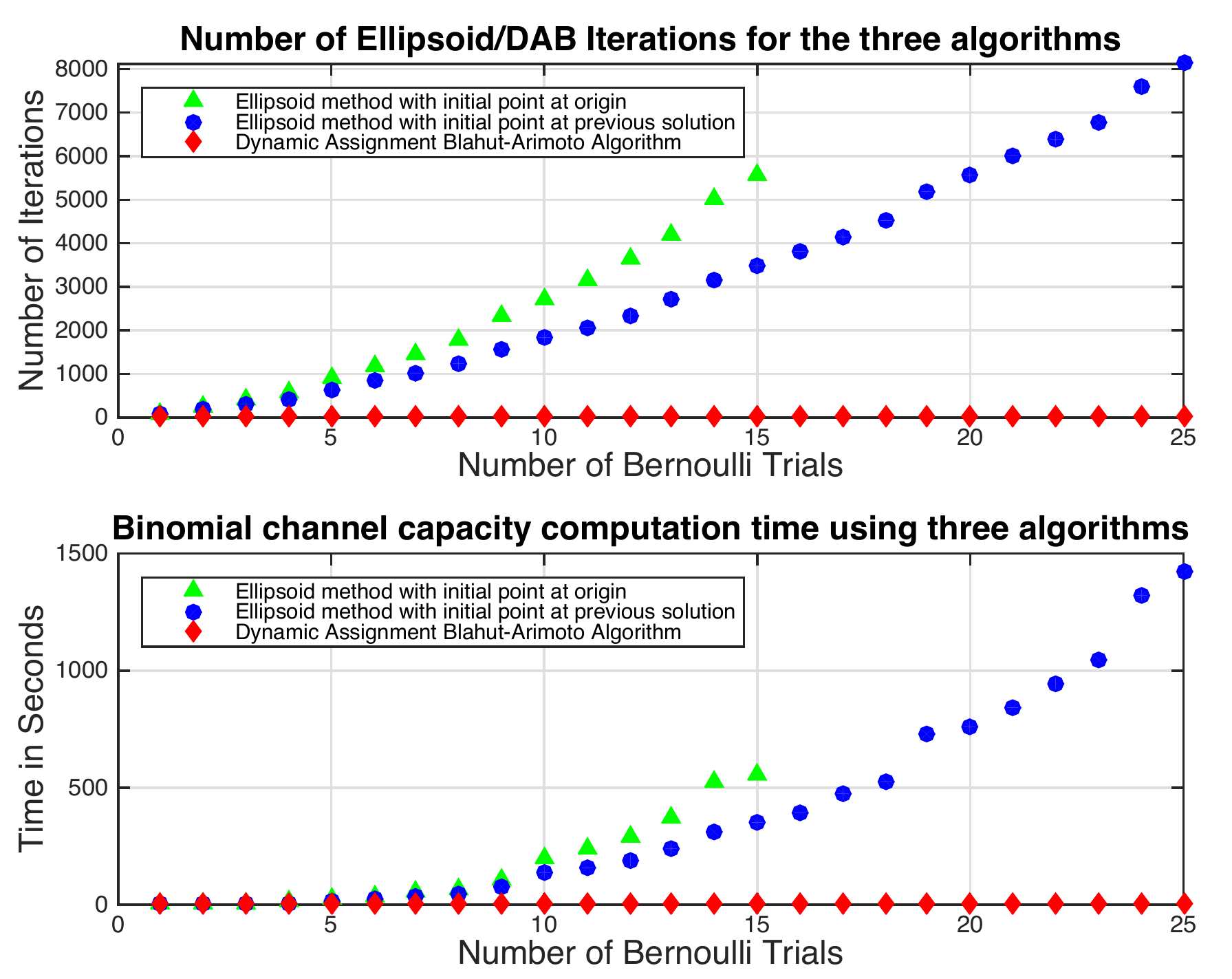}
\caption{Number of iterations and computation time in seconds (on a 2.5 GHz MacBook Pro purchased in 2014) running the Ellipsoid and Dynamic Assignment Blahut-Arimoto (DAB) algorithms implemented in Matlab to compute the binomial channel capacity: Ellipsoid method with initial point at the origin (green triangles), Ellipsoid method with initial point at position indicated by the solution for the previous $n$ (blue dots), and the DAB algorithm with initialization as described in Section \ref{sec:initialization} (red diamonds).}
\label{Figure:times_ellipsoid}
\end{figure}

However, there is a difficulty in knowing what the squared distance is before the problem has been solved.  This problem is avoided by selecting the initial $z$ vector for the $(n+1)^{st}$ case by using the solution obtained for the $n^\text{th}$~case as follows:
\begin{align*}
q_y^{(n+1,\text{initial})} &= \sum_{x \in A} P^{(n)}_{Y|X}(y|x) P^{(n,*)}_X(x) ~~ \forall y \in {0, \ldots, n},\\
z_y^{(n+1,\text{initial})} &= -\log e q_y^{(n+1,\text{initial})} \, .
\end{align*}
In this case, as shown in Figure  \ref{Figure:distances}, $P_0$ can often be the unscaled identity (or the identity scaled by a value less than one).   With such a close starting value, one would expect that the ellipsoid method would converge much more quickly.  However, as shown in Figure \ref{Figure:times_ellipsoid}, initializing the $z$ vector to the previously optimal point does not significantly improve performance, highlighting the slow convergence of the ellipsoid algorithm even when initialized to a favorable point. 

\section{Dynamic Assignment Blahut-Arimoto}
\label{sec:DAB}

This section introduces the Dynamic Assignment Blahut-Arimoto (DAB) algorithm, which computes the capacity and associated capacity-achieving distribution when the input alphabet is continuous but a capacity-achieving (or capacity-approaching) distribution is known to have finite support.

The DAB approach alternates between a Blahut-Arimoto step that optimizes the allocation of probability to a fixed set of mass points and a second step that adjusts the placement (and possibly the number) of mass points given the PMF identified by the previous Blahut-Arimoto step.

Algorithm 1 summarizes DAB.  This iterative algorithm starts by initializing the number of mass points $N^{(1)}$ and their locations $\mathcal{\tilde{X}}^{(1)}$.  DAB can increase the number of mass points if necessary.
During the $k^{\text{th}}$~iteration of the algorithm, first $\mathcal{\tilde{X}}^{(k)}$ is used with Blahut-Arimoto algorithm to maximize MI and find the corresponding maximizing distribution $\mathcal{\tilde{P}}^{(k)}$. This provides a lower bound on capacity.  Then, the distribution $p(y)$ induced by $\mathcal{\tilde{X}}^{(k)}$ and $\mathcal{\tilde{P}}^{(k)}$ are used to compute an upper bound on capacity. If these bounds are within a specified threshold, then DAB has found the capacity (to a specified level of precision) and the associated capacity-achieving (or capacity-approaching) input distribution. Otherwise the location of the mass points and/or the number of mass points needs to be updated. These updates occur in steps 4, 5, and 6 of Algorithm \ref{Alg:ROVAr}. The subsections that follow explore specific steps of Algorithm 1 in more detail.

	\begin{algorithm}[t]
	\caption{Dynamic Assignment Blahut Arimoto} \label{Alg:ROVAr}
{\bf Initialization:} 
	Select $\mathcal{\tilde{X}}^{(1)} = \begin{bmatrix}x_1& x_2&\ldots &x_{N^{(1)}} \end{bmatrix}$, the vector  specifying the $N^{(1)}$ mass point locations in increasing order.  
	Select the tolerance $\epsilon$ controlling the accuracy of the final capacity.  
	Set $k=1$.
	
	{\bf Iterations:}  
	Determination of the optimal $\mathcal{\tilde{X}}^*$, $\mathcal{\tilde{P}}^*$, and the capacity $C$ (within $\epsilon$ bits) proceeds as follows:
	\begin{enumerate}
	\item Given $\mathcal{\tilde{X}}^{(k)}$, use Blahut-Arimoto to compute the MI-maximizing distribution $\mathcal{\tilde{P}}^{(k)}$ and the corresponding MI $I^{(k)}$, which is a lower bound on $C$.
	\item Use the distribution $p(y)$ induced by $\mathcal{\tilde{X}}^{(k)}$ and $\mathcal{\tilde{P}}^{(k)}$ to compute the capacity upper bound $$D_{\text{max}}^{(k)} = \max_{x \in {\cal X}} D \bigl (p(y|x)\,\|\,p(y) \bigr ).$$
	\item If $D_{\text{max}}^{(k)} -I^{(k)} < \epsilon$ conclude by reporting  $\mathcal{\tilde{X}}^*=\mathcal{\tilde{X}}^{(k)}$, $\mathcal{\tilde{P}}^*=\mathcal{\tilde{P}}^{(k)}$, and $C= I^{(k)}$. Otherwise continue.
	\item Determine whether $N^{(k+1)} = N^{(k)}$ or $N^{(k)}+1$ and if the latter, update $\mathcal{\tilde{X}}^{(k+1)}$ to include the additional location.
	\item Determine direction vector $\mathcal{\tilde{D}}^{k}$ to adjust $\mathcal{\tilde{X}}$.
	\item Compute 
	\begin{equation}
	\label{eq:X+lambdaD}
	    \mathcal{\tilde{X}}^{(k+1)} = \mathcal{\tilde{X}}^{(k)}+ \lambda^* \mathcal{\tilde{D}}^{k} \, ,
	\end{equation}
	where
	\begin{equation}
	    \lambda^* = \arg \max_{\lambda} I\left(\mathcal{\tilde{X}}^{(k)}+ \lambda \mathcal{\tilde{D}}^{k}, \mathcal{\tilde{P}}^{(k)} \right) \, ,
	\end{equation}
	and $I(\mathcal{\tilde{X}},\mathcal{\tilde{P}})$ is the mutual information that results from an input distribution with mass points whose locations are described by the vector $\mathcal{\tilde{X}}$ with corresponding probabilities are described by the vector $\mathcal{\tilde{P}}$.
	\item $k=k+1$.
	\item Go to 1.
	\end{enumerate}
	\end{algorithm}
	
\subsection{Initialization}
\label{sec:initialization}
Unlike the Ellipsoid method, proper initialization can dramatically reduce computation time for DAB.  Consider again Figure \ref{Figure:MassPoints}.  The mass points often move only slightly as $n$ progresses; significant changes occur only when the mass point at $x=1/2$ splits or when a new mass point is born at $x=\frac{1}{2}$. Notice that the gentle evolution of the capacity-achieving distribution as a function of $n$ is such that the number of mass points never increases by more than one. 

DAB allows this behavior to be exploited.  A clever approach to initialization uses the previously computed capacity-achieving distribution of a channel in the same ``family'' with a slightly lower capacity as a starting point for its optimization.  For the binomial channel, the starting distribution $\mathcal{\tilde{X}}^{(1)} $ may be selected  as the capacity-achieving distribution for the binomial channel with one fewer Bernoulli trial.
Figure \ref{Figure:times_ellipsoid} shows how this initialization approach (and the general efficiency of DAB) lead to dramatically smaller values for run times and required iterations as compared to the Ellipsoid method.

\subsection{Upper Bound via Csiszar's  Min-Max Capacity Theorem}
Step 2 of DAB relies on  
Csiszar's Min-Max Capacity Theorem \cite{CsiszarInformtionTheoryBook1981}, which  states:
\begin{equation}
C = \min_{p(y) \in \{P_Y\}} \max_x D \bigl (p(y|x)\,\|\,p(y) \bigr )\, , \label{eq:Csiszar}
\end{equation}
where $\{P_Y\}$ is the set of distributions on $Y$ that can be induced by a valid input distribution. In fact, we can restate the dual problem found in Section \ref{sec:convex_optimization} in terms of Csiszar's Min-Max Capacity Theorem as follows:
\begin{equation}
\min_{p(y)}\left \{ \sum_{y} P_Y(y) -1 + \max_x D \bigl (p(y|x)\,\|\,p(y) \bigr ) \right \} \, .
\end{equation}
An upper bound on capacity follows directly from \eqref{eq:Csiszar}:  For any valid output distribution on ${\cal Y}$,
\begin{equation}
\label{eq:D-bound}
C \le \max_x D \left ( P_{Y|X=x}\|P_Y \right ) \, .
\end{equation}

\subsection{Determining direction vector $\mathcal{\tilde{D}}^{k}$}
\label{sec:direction}
In this step of DAB (step 5 of Alg. 1), a direction is selected along which $\mathcal{\tilde{X}}^k$ will be varied in step 6 to increase the mutual information $I(\mathcal{\tilde{X}},\mathcal{\tilde{P}})$.  This paper considers two approaches to selecting the direction: moving a single mass point (or symmetric mass-point pair)  and moving all points by setting the direction to be the relevant gradient.
\subsubsection{Moving a single mass point (or a symmetric pair)}
The technique considered in the original DAB  \cite{richard_d._wesel_efficient_2018} was to select and move a single mass point so that $\mathcal{\tilde{D}}^{k}= e_j$, which the vector with all elements set to zero except the \jth~element, which is set to one.  When the capacity-achieving distribution is known to have symmetry about its center, $\mathcal{\tilde{D}}^{k}$ should be selected as a symmetric pair of mass points so that 
\begin{equation}
    \mathcal{\tilde{D}}^{k}= e_j + e_{ N^{(k)} +1-j}.
\end{equation}
The binomial channel is an example where $\mathcal{\tilde{X}}^*$ is symmetric (about 1/2) so that a symmetric pair of mass point locations are set to 1 to form $\mathcal{\tilde{D}}^{k}$. The PIC does not have such symmetry.

{\em 1a) Proximity to $x_{\text{max}}$:}
In \cite{richard_d._wesel_efficient_2018}, motivated by reducing the upper bound of \eqref{eq:D-bound}, for the specific case of the binomial channel, the original DAB sets $\mathcal{\tilde{D}}^{k}= e_j$ where the mass point $x_j$ is the point in the interval bounded by $x_{\text{max}}$ and $1/2$ that is closest to $x_{\text{max}}$, where
\begin{align}
x_{\text{max}}^{(k)}  &= \arg \max_{x}  D \bigl (p(y|x)\,\|\,p(y) \bigr ) \,.
\end{align}

{\em 1b) Maximum derivative:}
\label{sec:MaxDerivative}
In this paper we also consider the alternative where DAB selects the mass point $x_j$ to maximizes the partial derivative
\begin{equation}
\label{eq:dIdx}
    \frac{\partial I\bigl(\mathcal{\tilde{X}}^{(k)}, \mathcal{\tilde{P}}^{(k)}\bigr)}{\partial x_j} = p_j 
     \frac{\partial D \bigl (p(y|x_j)\,\|\,p(y) \bigr )}{\partial x_j}\, .
\end{equation}

{\em 1c) Round Robin:} As a third alternative DAB could select the mass points (or symmetric pairs) in a round robin fashion.  This is a conservative approach that makes sure that every pair gets a chance to adjust its position.

\subsubsection{Moving along the gradient}
\label{sec:gradient}
\begin{figure}
\centering\includegraphics[width=22pc]{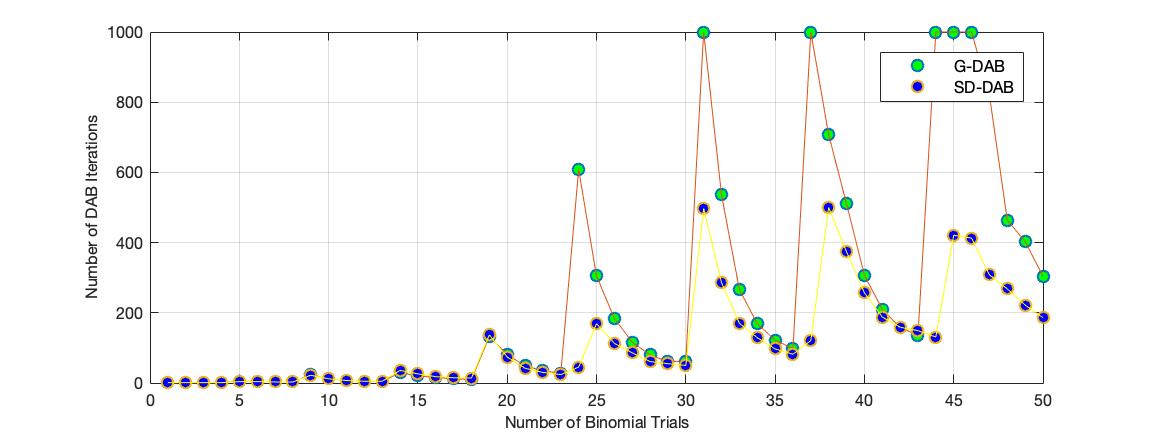}
\centering\includegraphics[width=22pc]{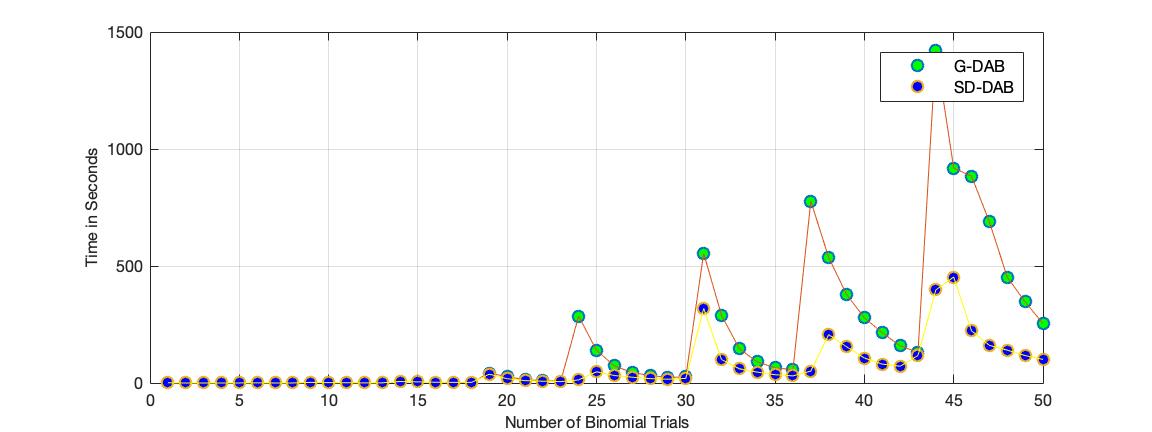}
\caption{Number of iterations and computation time in seconds (on a 2.5 GHz MacBook Pro purchased in 2018) running the Dynamic Assignment Blahut-Arimoto (DAB) algorithms implemented in Matlab to compute the binomial channel capacity with the single-derivative (SD) DAB described in Sec. \ref{sec:MaxDerivative} and the Gradient (G) DAB described in Sec. \ref{sec:gradient}  Overall, SD-DAB proves to be superior to G-DAB in computational speed and number of iterations.}
\label{Figure:times}
\end{figure}

Considering \eqref{eq:dIdx}, rather than selecting any single mass point $\mathcal{\tilde{D}}^{k}= e_j$, this approach sets $\mathcal{\tilde{D}}^{k}$ to the gradient:
 \begin{equation}
 \mathcal{\tilde{D}}^{k}=
        \nabla_\mathcal{\tilde{X}} I^{(k)}= \begin{bmatrix}
            \frac{\partial I\left(\mathcal{\tilde{X}}^{(k)}, \mathcal{\tilde{P}}^{(k)}\right)}{\partial x_1}\\
            \frac{\partial I\left(\mathcal{\tilde{X}}^{(k)}, \mathcal{\tilde{P}}^{(k)}\right)}{\partial x_2}\\
            \vdots\\
            \frac{\partial I\left(\mathcal{\tilde{X}}^{(k)}, \mathcal{\tilde{P}}^{(k)}\right)}{\partial x_N}
        \end{bmatrix} \, .
\end{equation}
When optimizing along the direction of the gradient, the value of scalar $\lambda$ in \eqref{eq:X+lambdaD} is limited so that the mass point locations cannot cross each other.  Figure~\ref{Figure:times} compares values for the running time and the number of iterations required to calculate the capacity and the optimal input distribution for the Binomial channel in Figure~\ref{Figure:MassPoints} based on methods 1b) and 2.  As can be seen moving a single mass point tends to result in faster convergence. Therefore, in the rest of the paper we will use this method for numerical evaluations.

\subsection{Determining $\lambda$ to maximize $I(\mathcal{\tilde{X}},\mathcal{\tilde{P}})$}
\label{sec:Lambda}
A line search routine (such as \texttt{fminbd} in Matlab) determines the value of $\lambda$ that maximizes $I(\mathcal{\tilde{X}},\mathcal{\tilde{P}})$ in step 6 of Algorithm~1.  Since mutual information is a concave function of the input distribution \cite{cover-book}, this line search is guaranteed to find the globally optimal point.  
As the number of mass points grows, fully optimizing $\lambda$ to maximize $I(\mathcal{\tilde{X}},\mathcal{\tilde{P}})$ provides more consistent performance than the approach in \cite{richard_d._wesel_efficient_2018} of moving the mass point in the direction of $x_{\text{max}}$ using a step size $\delta$.

\subsection{Determining when to increment $N^{(k)}$}
\label{sec:splitting}
There are three possible approaches to deciding when to increment $N^{(k)}$ in step 4 of Algorithm \ref{Alg:ROVAr}. 
\subsubsection{"Missing mass point" approach}
\label{sec:proximity} When applying approach 1a of Sec \ref{sec:direction} for the binomial channel, an additional mass point is added whenever none of the mass points in $\mathcal{\tilde{X}}^{(k)}$ lies in the interval bounded by $x_{\text{max}}$ and $1/2$.  If $N^{(k)}$ is even, a new point is added at $1/2$.  If  $N^{(k)}$ is odd, a new point is added by splitting the point at $1/2$ into two points that will then be pulled away from $1/2$ by the line search of step 7.

\subsubsection{Minimum derivative test}
\label{sec:minimumdIdx} When applying approaches 1b or 2 of Sec \ref{sec:direction}, additional mass point is added whenever the largest derivative is small enough that further improvement requires an additional mass point. The test in step 4 reveals when further mutual information increase is possible.  When the derivatives all become too small to allow the potential improvement identified in step 4, this is a clear indication  that an additional mass point is needed.  If  $N^{(k)}$ is odd, a new point is added by splitting the central point.  If $N^{(k)}$ is even, a new point is added in between the two central points.

\subsubsection{Negligible rate of change of $I^{(k)}$}
\label{sec:change_rate} When applying any of the approaches of Sec. \ref{sec:direction}, simply tracking the increase in mutual information can be an effective way to determine when an additional mass point is needed.  If the $I^{(k)}-I^{(k-1)}$ becomes negligible, then a new mass point is added.  As above, if  $N^{(k)}$ is odd, a new point is added by splitting the central point.  If $N^{(k)}$ is even, a new point is added in between the two central points.
\label{sec:change_rate}

\section{Numerical Evaluations}

This section uses  DAB to evaluate the capacity and the optimal input distribution for several instances of the PIC.  The assumed mechanism for particle transport is diffusion with coefficient $d$ from a point-source transmitter to the surface of a spherical receiver with radius $r$.  Let $\ell$ be the shortest distance between the point source and the receiver surface. Under this model, the motion of each particle can be represented using a random Brownian path in 3D space. 

Since we assume that the particles are either detected when they arrive at the receiver or they are never detectable, the time of arrival is given by the first time the particle reaches the receiver. For Brownian motion in 3D space, the first arrival time $T$ to the spherical receiver is a scaled L\'evy-distributed random variable where the scale constant is $\eta=\tfrac{r}{\ell+r}$ \cite{yilmaz20143dChannelCF}. This means that there is a non-zero probability that a particle never arrives at the receiver. Note that for Brownian motion in 1D space, $\eta=1$.  Using the iCDF of the scaled \levy distribution, we obtain
	\begin{align}
	\label{eq:tauDiff}
	\tau = F_T^{-1}(\rho) =\frac{c}{2\erfcinv^2(\rho/\eta)}.		
	\end{align}
	where $c=\tfrac{\ell^2}{2d}$ and $\erfcinv(.)$ is the inverse of the complementary error function $\erfc(.)$. We call a channel that relies on this diffusive transport the {\em diffusion-based PIC (DBPIC)}.


	\label{sec:results}

		\begin{figure}
		\begin{center}
			\includegraphics[width=0.95\columnwidth,keepaspectratio]{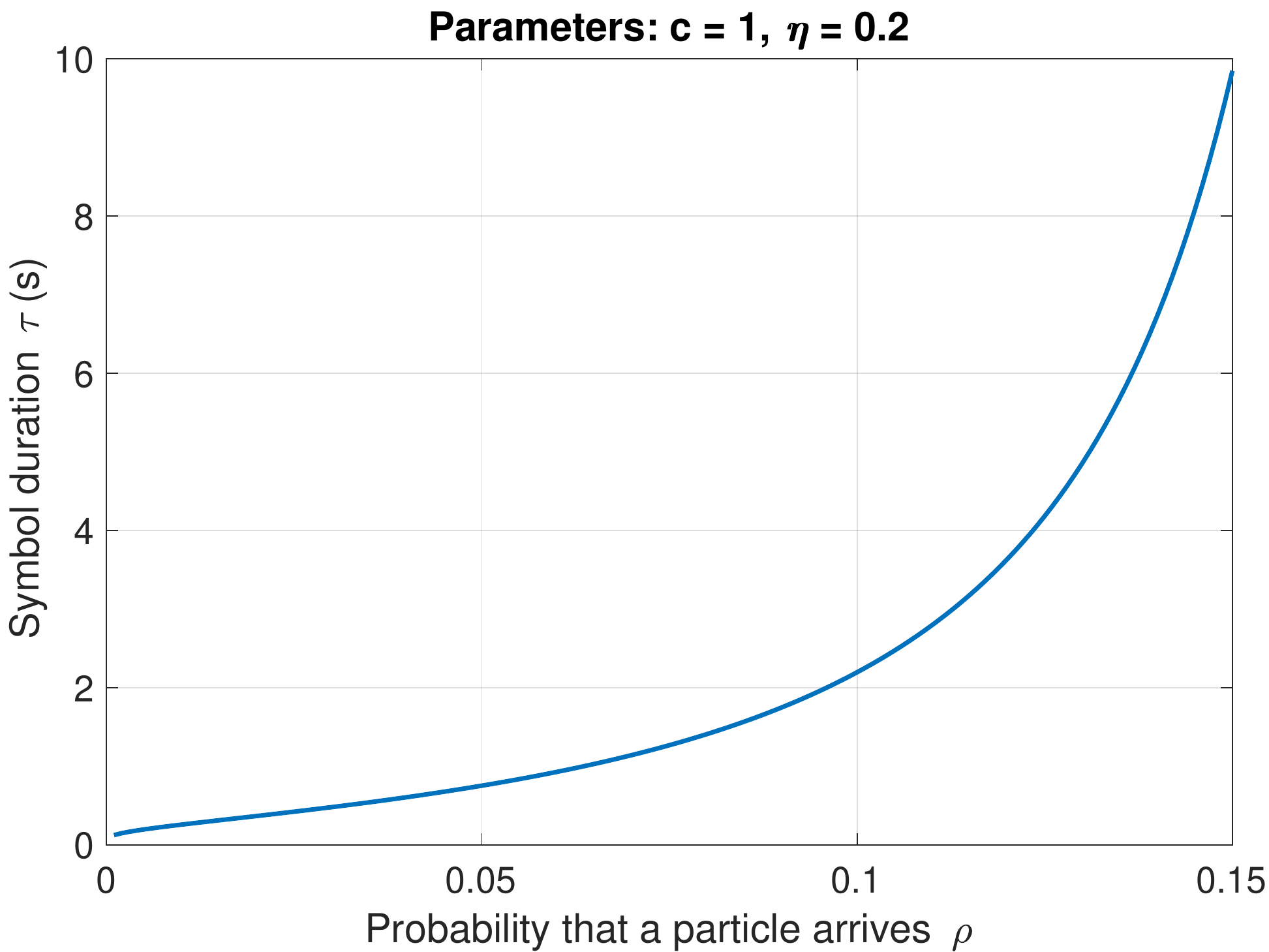}
		\end{center}
		\caption{\label{fig:iCDF} Plot of the iCDF of the $\mathcal{\tilde{X}}$ L\'evy-distribution.}
	\end{figure}

		\begin{figure}
	\begin{center}
		\includegraphics[width=0.95\columnwidth,keepaspectratio]{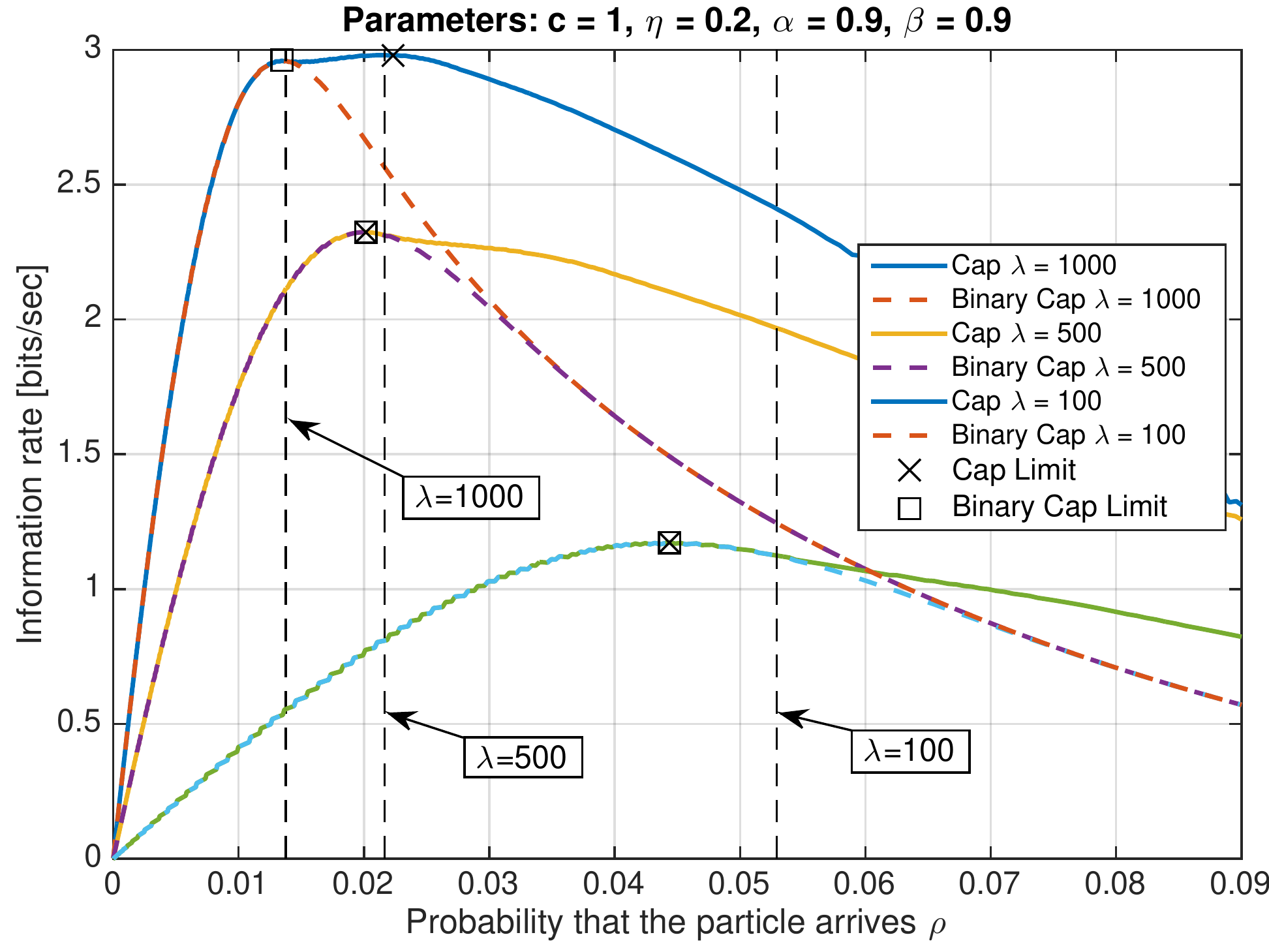}
	\end{center}
	\caption{\label{fig:capPlot} The information rate for binary-input (dashed lines) and the information rate with the optimal input based (solid lines) for different $\lambda$. The three vertical dashed lines indicate the $\rho$ value after which $m_\rho \theta_{\rho}$ in Proposition \ref{prop:CondiBinary} is greater than 3.3676.}
\end{figure}

		\begin{rem}
		Substituting \eqref{eq:tauDiff} into \eqref{eq:CapPIasP}, we observe that the diffusion coefficient $d$ has no effect on the optimal input distribution and the optimal $\rho$. This is despite the fact that the capacity increases linearly with $d$. This means that if the type of particle is changed, so long as the distance between the transmitter and the receiver is the same, and the receiver has the same radius, the optimal distribution and the optimal $\rho$ values will remain the same. Note that the change in capacity is due to the fact that a shorter or a longer symbol duration is required to achieve the same $\rho$ value.
	\end{rem}
	
	\begin{rem}
		If we consider a 1D environment\footnote{Note that a 1D environment is a good approximation if the system is confined inside a very narrow and long physical channel.} (i.e., $\eta=1$), we observe that the capacity decreases as $\tfrac{1}{l^2}$, and the distance $l$ does not affect the optimal input distribution and the optimal $\rho$. For a 3D environment however, changing the distance $l$ and the radius $r$ could affect the optimal $\rho$ and $P(x)$ values through $\eta$. 
	\end{rem}

	Figure~\ref{fig:iCDF} shows the iCDF  of the scaled \levy distribution in \eqref{eq:tauDiff} for $c=1$ and $\eta=0.2$. Note that a small increase in $\rho$ can require a large increase in symbol duration.   Therefore, larger $\rho$ values may not necessarily result in higher information rate in bits per second.	
	This effect is verified in Figure~\ref{fig:capPlot} where the information rate is plotted for three different particle generation rates, $\lambda$. The scaling factor for the \levy distribution is $\eta=0.2$, which means the distance between the transmitter and the receiver is four times the radius of the receiver. 
	
	The square markers indicate the maximum value of $\mathsf{C}^b(\rho)$ of \eqref{eq:binCapCEMperP}
	for the binary input distribution, and the $\times$-markers indicated $\mathsf{C}^*$ in \eqref{eq:capPI} for the optimal input distribution. For the case of $\lambda=1000$, Figure~\ref{fig:capPlot} shows that the binary input distribution does not maximize $\mathsf{C}^*$. The three vertical dashed lines indicate, for each choice of $\lambda$, the $\rho$ value after which $m_\rho \theta_{\rho}$ in Proposition \ref{prop:CondiBinary} is greater than 3.3676. We observe that for the $\rho$ values smaller than this critical value, the binary input is the optimal input distribution. 

\begin{figure}
	\centering\includegraphics[width=21pc]{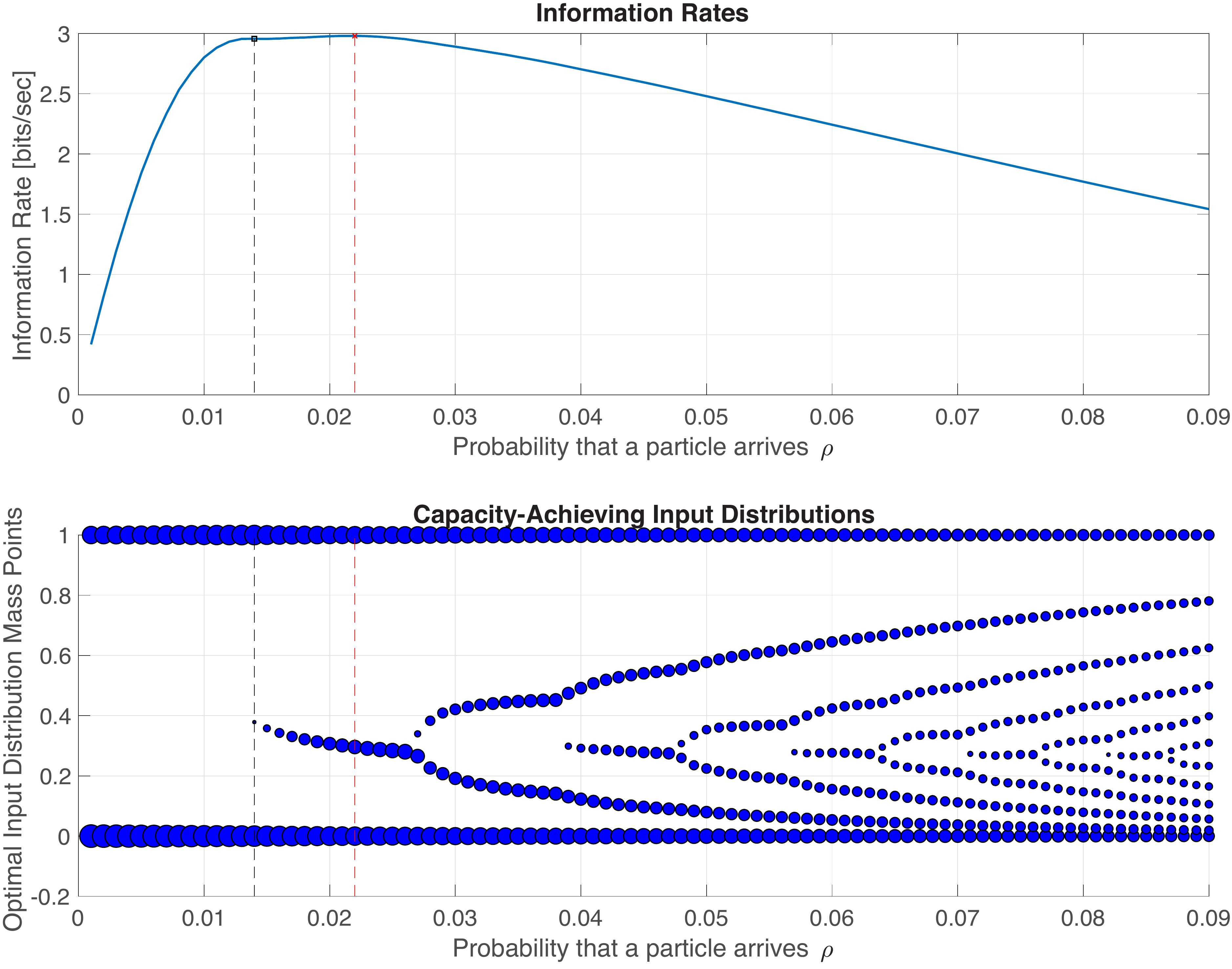}
	\caption{Information rates and capacity-achieving input distributions as a function of $\rho$ for the PIC with  $c=1, \eta = 0.2, \alpha = 0.9, \beta=0.9$ and $\lambda=1000$. Finite-support capacity-achieving distributions were obtained by the algorithm described in Section \ref{sec:DAB}. To generate the results of this graph, we use methods described in \ref{sec:MaxDerivative}b and \ref{sec:change_rate}. \label{fig:plot_of_MtoIRMP}}
\end{figure}

\begin{figure}
	\centering\includegraphics[width=21pc]{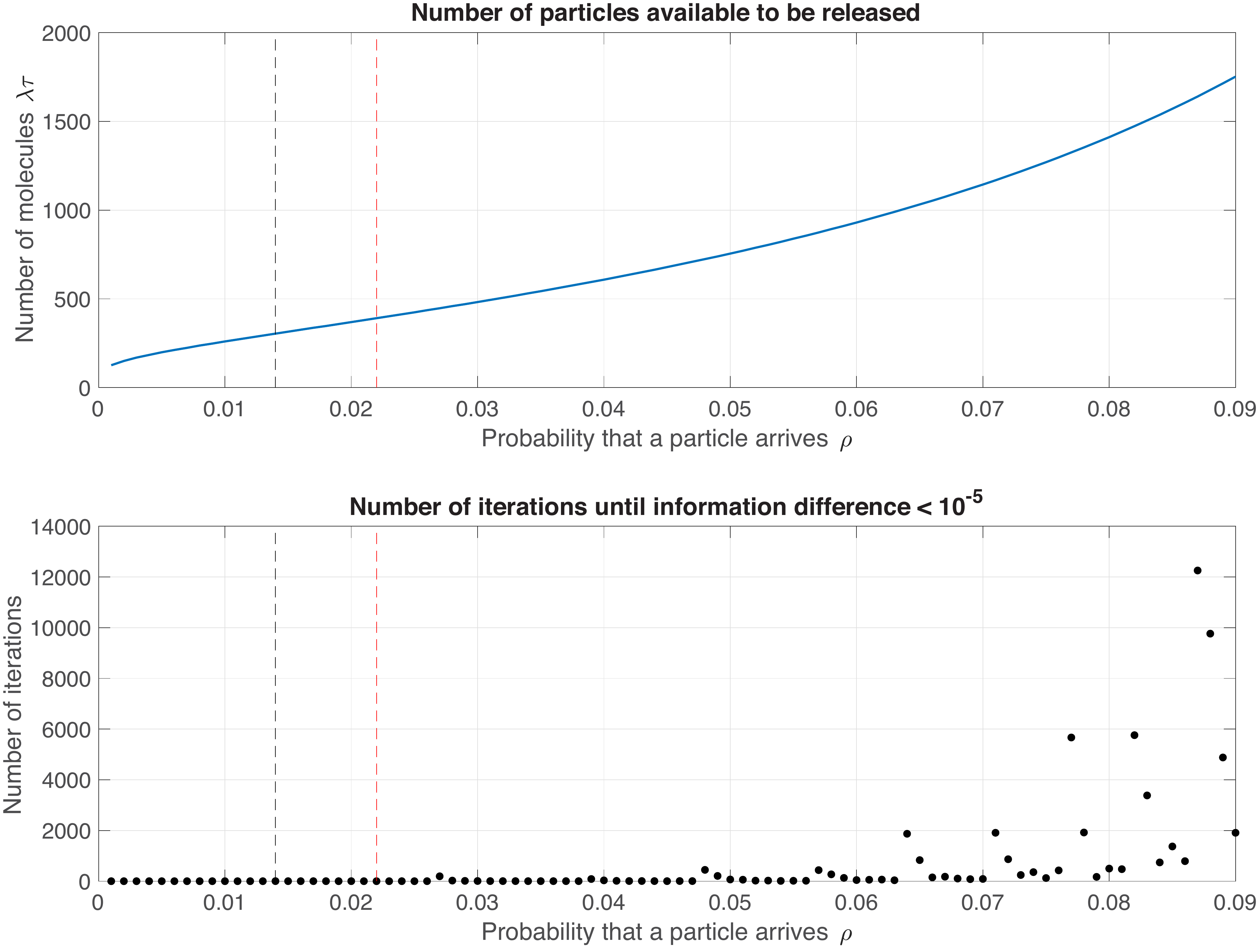}
	\caption{Number of available particles vs. $\rho$ and number of DAB iterations until the  information difference $D \left ( P_{Y|X=x}\|P_Y \right ) - I(X;Y)$  is below $10^{-5}$ as a function of $\rho$ for the PIC with  the same parameters and algorithm as in Figure \ref{fig:plot_of_MtoIRMP}.}
	\label{fig:plot_of_Mto_n_and_iter}
\end{figure}

DAB provides the capacity and optimal mass point locations for a sequence of $\rho$  values for the PIC channel.  Unlike the binomial channel, there is no longer any assumption of symmetry in the input distribution. Therefore, the new mass points are introduced at a location between the two middle mass points, or the middle mass point is split. Figures \ref{fig:plot_of_MtoIRMP} and \ref{fig:plot_of_Mto_n_and_iter} summarize the results of application of DAB to the PIC with parameters $c=1, \eta = 0.2, \alpha = 0.9, \beta=0.9$ and $\lambda=1000$. Particularly, the top plot in Figure \ref{fig:plot_of_MtoIRMP} shows the channel capacity, and the bottom plot shows the capacity achieving distribution corresponding to the each $\rho$ value. The black dashed line indicates the first $\rho$ value for which binary distribution is no longer capacity achieving, and the red dashed line indicates the location of the $\rho$ value that achieves the largest information rate. Again we can verify that Proposition \ref{prop:CondiBinary} holds, and that only 3 mass points are required to achieve the highest information rate.

\begin{figure}
	\centering\includegraphics[width=21pc]{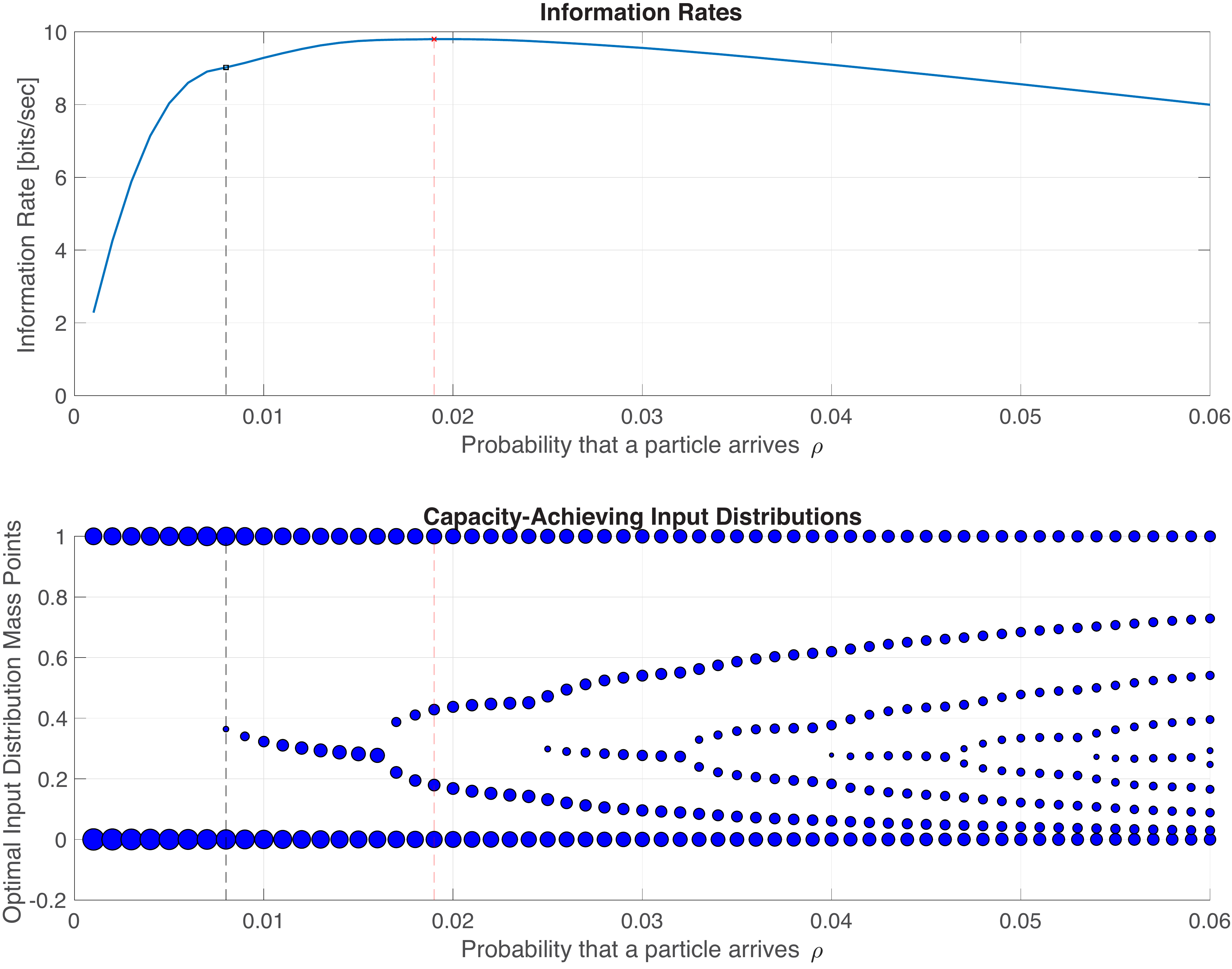}
	\caption{Information rates and capacity-achieving input distributions as a function of $\rho$ for the PIC with  $c=0.5, \eta = 0.3, \alpha = 0.95, \beta=0.95$ and $\lambda=5000$. Finite-support capacity-achieving distributions were obtained by the algorithm described in Section \ref{sec:DAB}. To generate the results of this graph, we use methods described in \ref{sec:MaxDerivative}b and \ref{sec:change_rate}. \label{fig:plot_of_MtoIRMP2}}
\end{figure}

The top of Figure~\ref{fig:plot_of_Mto_n_and_iter} plots the number of particles that can be released by the transmitter as a function of $\rho$. The bottom shows the number of iterations required until  $D \left ( P_{Y|X=x}\|P_Y \right ) - I(X;Y)$  is below $10^{-5}$. We observe that even with multiple mass points, the number of iterations can be quite small when no new mass point has been recently introduced.  As can be seen in Fig. \ref{fig:plot_of_MtoIRMP}, in these cases the new mass point positions are small modifications to the previous mass point positions, which were used to initialize DAB. However, as the number of mass points increases the maximum number of iterations required by DAB (typically right after a mass point is introduced) does increase. 

To demonstrate that the input distribution that achieves capacity can have more than 3 mass points, in Figure~\ref{fig:plot_of_MtoIRMP2}, we consider a system with $c=0.5, \eta = 0.3, \alpha = 0.95, \beta=0.95$ and $\lambda=5000$. Recall that the black dashed line indicates the first $\rho$ value for which binary distribution is no longer capacity achieving, and the red dashed line indicates the location of the $\rho$ value that achieves the largest information rate. As can be seen, for this system, the largest information rate is achieved when the number of mass points is equal to 4. Again the results verify that Proposition \ref{prop:CondiBinary} holds.

	\section{Conclusions}
	\label{sec:conclusion}
	\vspace{-0.15cm}
	This paper introduces the PIC and analyzes its capacity and the associated capacity-achieving distribution.  We show that the optimal input distribution for this channel always has mass points at probabilities of release having values of zero and one.  For diffusion-based propagation, the diffusion coefficient, and hence the type of the particles used, does not affect the optimal input distribution. We then derived capacity for the binary input diffusion-based PIC and present conditions under which a binary input is optimal for this channel. 
	
	This paper also introduces DAB as an efficient algorithm for finding the capacity and associated capacity-achieving distribution when the input alphabet is continuous but a capacity-achieving (or capacity-approaching) distribution is known to have finite support. This paper applies DAB to the binomial channel and the PIC.
	DAB provides numerical results illustrating that binary input is optimal for systems where the transmitter cannot generate particles at rates that satisfy Proposition \ref{prop:CondiBinary}. This can be thought of as the low SNR regime.  Future work on the PIC could explore the effect of ISI.  Future applications of DAB include a wide range of channels including peak and power limited additive white Gaussian noise channels and optical channels.

	\bibliographystyle{IEEEtran}
	\bibliography{IEEEabrv,MolCom,BinomialChannel}

\begin{thebibliography}{10}
\providecommand{\url}[1]{#1}
\csname url@samestyle\endcsname
\providecommand{\newblock}{\relax}
\providecommand{\bibinfo}[2]{#2}
\providecommand{\BIBentrySTDinterwordspacing}{\spaceskip=0pt\relax}
\providecommand{\BIBentryALTinterwordstretchfactor}{4}
\providecommand{\BIBentryALTinterwordspacing}{\spaceskip=\fontdimen2\font plus
\BIBentryALTinterwordstretchfactor\fontdimen3\font minus
  \fontdimen4\font\relax}
\providecommand{\BIBforeignlanguage}[2]{{%
\expandafter\ifx\csname l@#1\endcsname\relax
\typeout{** WARNING: IEEEtran.bst: No hyphenation pattern has been}%
\typeout{** loaded for the language `#1'. Using the pattern for}%
\typeout{** the default language instead.}%
\else
\language=\csname l@#1\endcsname
\fi
#2}}
\providecommand{\BIBdecl}{\relax}
\BIBdecl

\bibitem{ParticleIntensityModulation}
N.~{Farsad} \emph{et~al.}, ``Capacity of molecular channels with imperfect
  particle-intensity modulation and detection,'' in \emph{IEEE International
  Symposium on Information Theory (ISIT)}, 2017, pp. 2468--2472.

\bibitem{richard_d._wesel_efficient_2018}
{Richard D. Wesel} \emph{et~al.}, ``Efficient {{Binomial Channel Capacity
  Computation}} with an {{Application}} to {{Molecular Communication}},'' in
  \emph{Proc. {{Inf}}. {{Theory}} and {{Applications}} ({{ITA}}) {{Workshop}}},
  La Jolla (CA), Feb. 2018.

\bibitem{far16ST}
N.~Farsad \emph{et~al.}, ``A comprehensive survey of recent advancements in
  molecular communication,'' \emph{{IEEE} Commun. Surveys Tuts.}, vol.~18,
  no.~3, pp. 1887--1919, 2016.

\bibitem{sri12}
K.~V. Srinivas \emph{et~al.}, ``Molecular communication in fluid media: The
  additive inverse gaussian noise channel,'' \emph{{IEEE} Trans. Inf. Theory},
  vol.~58, no.~7, pp. 4678--4692, 2012.

\bibitem{li14}
H.~Li \emph{et~al.}, ``Capacity of the memoryless additive inverse gaussian
  noise channel,'' \emph{{IEEE} J. Sel. Areas Commun.}, vol.~32, no.~12, pp.
  2315--2329, 2014.

\bibitem{rose2016inscribed}
C.~Rose and I.~S. Mian, ``Inscribed matter communication: Part i,'' \emph{IEEE
  Transactions on Molecular, Biological and Multi-Scale Communications},
  vol.~2, no.~2, pp. 209--227, 2016.

\bibitem{far18_MTCcap}
N.~{Farsad} \emph{et~al.}, ``Capacity limits of diffusion-based molecular
  timing channels with finite particle lifetime,'' \emph{IEEE Transactions on
  Molecular, Biological and Multi-Scale Communications}, vol.~4, no.~2, pp.
  88--106, 2018.

\bibitem{ein2011}
A.~Einolghozati \emph{et~al.}, ``Capacity of discrete molecular diffusion
  channels,'' in \emph{Proc. IEEE Int. Symp. on Inf. Theory}, 2011.

\bibitem{tah15}
M.~Tahmasbi and F.~Fekri, ``On the capacity achieving probability measures for
  molecular receivers,'' in \emph{Proc. IEEE Inf. Theory Workshop}, 2015.

\bibitem{ami15PtoP}
G.~Aminian \emph{et~al.}, ``On the capacity of point-to-point and
  multiple-access molecular communications with ligand-receptors,''
  \emph{{IEEE} Trans. Mol. Biol. Multi-Scale Commun.}, vol.~1, no.~4, pp.
  331--346, 2015.

\bibitem{gha15}
S.~Ghavami \emph{et~al.}, ``Information rates of ask-based molecular
  communication in fluid media,'' \emph{{IEEE} Trans. Mol. Biol. Multi-Scale
  Commun.}, vol.~1, no.~3, pp. 277--291, 2015.

\bibitem{ami15}
G.~Aminian \emph{et~al.}, ``Capacity of diffusion-based molecular communication
  networks over {LTI}-poisson channels,'' \emph{{IEEE} Trans. Mol. Biol.
  Multi-Scale Commun.}, vol.~1, no.~2, pp. 188--201, 2015.

\bibitem{Lu15}
Y.~Lu \emph{et~al.}, ``Comparison of channel coding schemes for molecular
  communications systems,'' \emph{{IEEE} Trans. Commun.}, vol.~63, no.~11, pp.
  3991--4001, Nov 2015.

\bibitem{guo16}
W.~Guo \emph{et~al.}, ``Molecular communications: channel model and physical
  layer techniques,'' \emph{Wireless Commun.}, vol.~23, no.~4, pp. 120--127,
  2016.

\bibitem{KomninakisISIT2001}
C.~Komninakis \emph{et~al.}, ``Capacity of the binomial channel, or minimax
  redundancy for memoryless sources,'' in \emph{Proceedings of the IEEE
  International Symposium on Information Theory}, June 2001.

\bibitem{ver69}
W.~Vervaat, ``Upper bounds for the distance in total variation between the
  binomial or negative binomial and the poisson distribution,''
  \emph{Statistica Neerlandica}, vol.~23, no.~1, pp. 79--86, 1969.

\bibitem{sha90}
S.~Shamai, ``Capacity of a pulse amplitude modulated direct detection photon
  channel,'' \emph{IEE Proceedings I - Communications, Speech and Vision}, vol.
  137, no.~6, pp. 424--430, Dec 1990.

\bibitem{lap98}
A.~Lapidoth and S.~Shamai, ``The poisson multiple-access channel,''
  \emph{{IEEE} Trans. Inf. Theory}, vol.~44, no.~2, pp. 488--501, Mar 1998.

\bibitem{cao14}
J.~Cao \emph{et~al.}, ``Capacity-achieving distributions for the discrete-time
  poisson channel--{Part I}: General properties and numerical techniques,''
  \emph{{IEEE} Trans. Commun.}, vol.~62, no.~1, pp. 194--202, 2014.

\bibitem{WitsenhausenIT1980}
H.~S. Witsenhausen, ``Some aspects of convexity useful in information theory,''
  \emph{IEEE Transactions on Information Theory}, vol.~26, no.~3, pp. 265--271,
  May 1980.

\bibitem{Dubins1962}
L.~E. Dubins, ``On extreme points of convex sets,'' \emph{Journal of
  Mathematical Analysis and Applications}, vol.~5, no.~2, pp. 237--244, 1962.

\bibitem{GallagerBook1968}
R.~G. Gallager, \emph{Informatin Theory and Reliable Communication}.\hskip 1em
  plus 0.5em minus 0.4em\relax New York: Wiley, 1968.

\bibitem{cover-book}
T.~M. Cover and J.~A. Thomas, \emph{Elements of Information Theory 2nd
  Edition}, 2nd~ed.\hskip 1em plus 0.5em minus 0.4em\relax
  {Wiley-Interscience}, 2006.

\bibitem{tal02}
L.~G. Tallini \emph{et~al.}, ``On the capacity and codes for the z-channel,''
  in \emph{Proc. IEEE Int. Symp. on Inf. Theory}, 2002.

\bibitem{non-uniform}
A.~A. Farid and S.~Hranilovic, ``Channel capacity and non-uniform signalling
  for free-space optical intensity channels,'' \emph{IEEE Journal on Selected
  Areas in Communications}, vol.~27, no.~9, pp. 1553--1563, December 2009.

\bibitem{poisson}
J.~Cao \emph{et~al.}, ``Capacity and nonuniform signaling for discrete-time
  poisson channels,'' \emph{IEEE/OSA Journal of Optical Communications and
  Networking}, vol.~5, no.~4, pp. 329--337, April 2013.

\bibitem{poisson_part_1}
------, ``Capacity-achieving distributions for the discrete-time poisson
  channelâ-part i: General properties and numerical techniques,''
  \emph{Communications, IEEE Transactions on}, vol.~62, pp. 194--202, 01 2014.

\bibitem{constrain}
J.~G. Smith, ``The information capacity of amplitude- and variance-constrained
  scalar gaussian channels,'' \emph{Information and Control}, vol.~18, pp.
  203--219, 1971.

\bibitem{char}
J.~Huang and S.~P. Meyn, ``Characterization and computation of optimal
  distributions for channel coding,'' \emph{IEEE Transactions on Information
  Theory}, vol.~51, no.~7, pp. 2336--2351, July 2005.

\bibitem{optical}
A.~Lapidoth \emph{et~al.}, ``On the capacity of free-space optical intensity
  channels,'' \emph{IEEE Transactions on Information Theory}, vol.~55, no.~10,
  pp. 4449--4461, Oct 2009.

\bibitem{photon}
S.~Shamai, ``On the capacity of a direct-detection photon channel with
  intertransition-constrained binary input,'' \emph{IEEE Transactions on
  Information Theory}, vol.~37, no.~6, pp. 1540--1550, Nov 1991.

\bibitem{poisson_2}
A.~Lapidoth and S.~M. Moser, ``On the capacity of the discrete-time poisson
  channel,'' \emph{IEEE Transactions on Information Theory}, vol.~55, no.~1,
  pp. 303--322, Jan 2009.

\bibitem{poisson_3}
J.~Cao \emph{et~al.}, ``Capacity-achieving distributions for the discrete-time
  poisson channel—part ii: Binary inputs,'' \emph{IEEE Transactions on
  Communications}, vol.~62, no.~1, pp. 203--213, January 2014.

\bibitem{CoverBook}
T.~M. Cover and J.~A. Thomas, \emph{Elements of Information Theory}.\hskip 1em
  plus 0.5em minus 0.4em\relax John Wiley \& Sons, 1991.

\bibitem{BlahutTIT1972}
R.~Blahut, ``Computation of channel capacity and rate-distortion functions,''
  \emph{IEEE Transactions on Information Theory}, vol.~18, no.~4, pp. 460--473,
  April 1972.

\bibitem{Luenberger}
D.~G. Luenberger, \emph{Optimization by Vector Space Methods}.\hskip 1em plus
  0.5em minus 0.4em\relax John Wiley \& Sons, Inc., 1969.

\bibitem{Khachiyan1979}
L.~G. Khachiyan, ``A polynomial algorithm in linear programming,''
  \emph{Doklady Akademiia Nauk SSSR (translated in Soviet Mathematics Doklady},
  vol.~20, pp. 191--194, 1979.

\bibitem{EllipsoidSurvey1981}
R.~G. Bland \emph{et~al.}, ``Feature article---the ellipsoid method: A
  survey,'' \emph{Operations Research}, vol.~29, no.~6, pp. 1039--1091,
  https://doi.org/10.1287/opre.29.6.1039 1981.

\bibitem{CsiszarInformtionTheoryBook1981}
I.~Csis{\'a}r and J.~K{\"o}rner, \emph{InformationTheory: Coding Theorems for
  Discrete Memoryless Systems}, ser. Probability and Mathematical Statistics,
  Z.~Birnbaum and E.~Lukacs, Eds.\hskip 1em plus 0.5em minus 0.4em\relax New
  York - San Francisco - London: Academic Press, 1981. See Theorem 3.4.

\bibitem{yilmaz20143dChannelCF}
H.~B. Yilmaz \emph{et~al.}, ``Three-dimensional channel characteristics for
  molecular communications with an absorbing receiver,'' \emph{{IEEE} Commun.
  Lett.}, vol.~18, no.~6, pp. 929--932, 2014.

\end{thebibliography}


\end{document}